\documentclass[trackchanges,twocolumn]{aastex701}

\usepackage{graphicx}
\usepackage{caption}
\usepackage[rightcaption]{sidecap}
\usepackage{wrapfig}
\usepackage{longtable}

\begin{document}
\title{Kepler-50: Two Planets in Close Resonance Perturbed by a Non-transiting Planet}
\correspondingauthor{Daniel Jontof-Hutter}
\email{djontofhutter@pacific.edu}
\affiliation{Department of Physics and Astronomy\\
University of the Pacific\\ 
601 Pacific Avenue\\
Stockton, CA 95211, USA}

\author[0000-0002-6227-7510]{Daniel Jontof-Hutter}
\affiliation{Department of Physics and Astronomy\\
University of the Pacific\\ 
601 Pacific Avenue\\
Stockton, CA 95211, USA}
\email{djontofhutter@pacific.edu}

\author[0000-0002-5904-1865]{Jason F. Rowe}
\affil{Department of Physics and Astronomy\\
Bishops University \\
2600 Rue College \\
Sherbrooke, QC J1M 1Z7,
Canada}
\email{jason.rowe@ubishops.ca}

\author[0000-0001-6513-1659]{Jack J. Lissauer}
\affiliation{Planetary Science \& Exobiology Division\\
MS 245-3\\
NASA Ames Research Center \\
 Moffett Field, CA 94035, USA}
\email{ Jack.Lissauer@nasa.gov }

\author[0000-0003-3750-0183]{Daniel C. Fabrycky}
\affiliation{Department of Astronomy and Astrophysics\\
University of Chicago\\
5640 South Ellis Avenue\\
Chicago, IL 60637, USA}
\email{fabrycky@uchicago.edu}

\begin{abstract}
Kepler-50 has two transiting planets close to the 6:5 commensurability. We analyze transit times measured from Kepler data with dynamical fits. We attribute the residual signal to a near-resonance non-transiting third planet in the system. The parameters of the third planet are degenerate. We identified 55 (15) regions of interest for an exterior (interior) perturber, and characterize system parameters in each region of interest with posterior sampling. We find that removing samples that are not long-term stable provides few additional constraints on posteriors, and that the transit times do not confirm the likely 2-body resonant state of the transiting planets as $\approx$ 68\% of posterior samples that are likely stable are also likely in libration. We identify a small fraction of samples that are in 3-body libration. Despite the degeneracies, we recover robust, tightly constrained masses for Kepler-50 b and c ($2.35^{+0.48}_{-0.43}$ $M_{\oplus}$  and $4.43^{+0.45}_{-0.49}$ $M_{\oplus}$, respectively) and strongly detected non-zero relative eccentricity for Kepler-50 b and c with ($e_c\cos\varpi_c$ -- $e_b\cos\varpi_b$) = --0.045 $\pm 0.003$, and ($e_c\sin\varpi_c$ -- $e_b\sin\varpi_b$) = --0.030 $\pm 0.004$. These close neighbors are in the radius valley and sub-Neptune by size, respectively ($1.750 \pm 0.038$ $R_{\oplus}$ and $2.079 \pm 0.060$ $R_{\oplus}$). With one of the closest planet pairs known and one of the closest to first-order commensurability, the Kepler-50 system is an especially valuable benchmark for planet formation theory.  
\end{abstract}

\section{Introduction}
Models of planetary system formation in a gaseous protoplanetary disk show that convergent migration naturally leads to resonant capture (\citealt{Lee2003,Goldreich2014,Afkanpour2024,Batygin2026a}). Low mass planets can form well outside of their ultimate orbits and migrate inwards over great distances if they remain in the Type I regime (\citealt{Terquem2007,Baruteau2008,Mustill2011}). This mechanism has been invoked to explain chains of resonant low-mass planets like TRAPPIST-1 (\citealt{Luger2017,Tamayo2017}), Kepler-223 \citep{Mills2016}, Kepler-80 \citep{Macdonald2016}, Kepler-60 \citep{Goz2016}, and K2-138 \citep{MacDonald2022}. 

The Kepler mission discovered more than 700 systems with multiple transiting planets \citep{Lissauer2024}. While there are several well-studied planetary systems with mean motion resonances (e.g., \citealt{Panichi2019,Goz2016,Mills2016}), the majority of planet pairs in the Kepler dataset are not near resonance (\citealt{Lissauer2011b,Fabrycky2014}), at odds with the migration model and suggesting that the assembly of resonances may not be the typical end-state of planetary system formation (\citealt{Chiang2013,Hansen2013}). 

To reconcile theory and observations, several studies invoke processes that break resonant chains post-formation (\citealt{Goldreich2014,Afkanpour2024,Dai2024,Hansen2024,Huang2025}).

However, challenges remain to explain systems with extremely proximate orbital period ratios, like Kepler-36 (\citealt{Deck2012, Batygin2026a}). Resonant trapping at high index mean motion resonance likely requires convergent migration that avoids capture in low-index resonances 2:1, 3:2, 4:3 etc before capture into resonance at small period ratio. 

Kepler-50 has two transiting $\sim 2 R_\oplus$ planets with one of the most proximate orbital period ratios discovered by Kepler. With period ratio $P_{c}/P_{b}$ = 1.20016 (close to 6:5), the pair ranks 9th among over 1000 adjacent planetary pairs in period ratio from the Kepler dataset \citep{Lissauer2024}. The pair ranks 2nd in proximity to first-order commensurability, given the transit periods recorded by the Kepler mission. A formation scenario that invokes convergent migration would likely require the planets to avoid permanent capture in lower index mean motion resonances (MMR) before being trapped in the 6:5 resonance. However, unlike Kepler-36, where the planets have an orbital period ratio $\frac{P'}{P} \approx 1.172$, the pair transiting Kepler-50 remain extremely close to a commensurate period ratio. \cite{Batygin2026a} found that convergent migration can enable passage through low-index MMRs and allow capture in the 7:6 resonance with reasonable constraints on the disk structure. In addition, in the absence of additional physics (e.g, braking migration at an inner edge \citealt{Pan2025}), convergence through low-index MMR may require the mass of the inner planet to be less than that of its neighbor. 

Kepler-50 provides a useful test of this scenario, although both planets are similar in size, and the constraints on their masses from radial velocity spectroscopy (RV) are uninformative \citep{Weiss2024}. Nevertheless, both planets exhibit strongly detected transit timing variations (TTVs) (\citealt{Steffen2012,Steffen2013a}), raising the prospect of precise mass measurements.

\cite{Jontof-Hutter2021} noted the poor fit of a two-planet model for the Kepler-50 TTVs. Several studies have explored the possibility of additional planets in systems with detected TTVs, which are poorly accounted for from the transiting planets alone (\citealt{Agol2005,Holman2005,Nesvorny2019}). 

While there are several known planets with TTVs that cannot be attributed to transiting neighboring planets, there have been few unique characterizations of the masses and orbits of non-transiting planets from the TTVs that they induce on transiting planets. Unless a TTV signal has high signal-to-noise ratio (S/N) (like Kepler-88 \cite{Nesvorny2013}), the solutions for a non-transiting perturber generally have degenerate orbital periods and phases (e.g., Kepler-82, Kepler-411, Kepler-725, KOI-134, Kepler-138, and TOI-4562 \citealt{Lammers2026}, Kepler-89 and Kepler-396 \citealt{Jontof-Hutter2022} and Kepler-51 \citealt{Masuda2024}). Unlike earlier studies that characterize one or a few potential solutions, \cite{Masuda2024} explores solutions for an additional planet orbiting Kepler-51 over a wide range of possible orbital periods. \cite{Jontof-Hutter2015b} noted that multiple possible solutions with an additional planet could improve the TTV models of Kepler-50. In this study, we explore such solutions methodically. We characterize the planetary masses and orbital parameters and assess their long term stability. 

In Section 2, we summarize our procedure to measure the transit times of the Kepler lightcurves. In Section 3 we explore two-planet fits to the measured transit times and justify an exploration of three-planet models. In Section 4 we perform an extensive grid search in dynamical mass and orbital parameter space for a third planet interior to Kepler-50 b and exterior to Kepler-50 c. In Section 5 we characterize the solutions found in Section 4 with posterior sampling. We consider whether the requirement of long-term stability or the requirement of libration in the 6:5 resonance among the posterior samples imposes useful constraints on all three planets. In Section 6, we summarize our results and characterize the masses of Kepler-50 b and c.  

\section{Light Curve Analysis}\label{sec:LightCurve}
We measured mid-transit times from the Kepler light curve using Fourier decomposition to remove stellar noise caused by semi-periodic behavior, such as spot modulation and stellar rotation.  We iteratively calculated a periodogram with the Lomb-Scargle algorithm, updating the model fit until a noise threshold is met. Our joint transit and Fourier model reduces colored noise by a factor of $\sim 2$ as the transit template has a more accurate out-of-transit baseline and exhibits fewer outliers. Hence, our transit timing measurements compare favorably to prior catalogs (\citealt{Rowe2015b,Holczer2016}). Nevertheless, a few outlying transit times remain, perhaps due to residual short-duration stellar activity such as flares. 

We measured 312 transit times for Kepler-50 b and c and list them in the Appendix in Table~\ref{tab:TT}. Given that our dynamical fits include the unknown component of an additional planet, we opted against excluding outliers from our TTV analysis with an arbitrary criterion, and revisit this issue in Section 4. 

\section{Two-planet models}\label{sec:2planets}
We ran two-planet fits using differential evolution Markov Chain Monte Carlo (MCMC). These models have 10 parameters with free orbital periods $P$, first transit after epoch $T_{0}$, eccentricity vector components  $h=e\sin\omega$ and $k=e\cos\omega$ (with Gaussian priors of width 0.02) and positive definite planetary masses. 

Given the weak sensitivity of TTV signals to moderate mutual inclinations \citep{Nesvorny2014}, we assumed coplanar orbits, with ascending nodes $\Omega = 0$. Since $\varpi = \Omega + \omega$ reduces to $\omega$ when $\Omega = 0$, the assumption of coplanar orbits means $\omega = \varpi$; we therefore adopt the longitude of pericenter $\varpi$ throughout the remainder of this paper.
 
Our chosen epoch is Barycentric Julian Date (BJD): 2454900+780. Our best-fit model was a poor fit $\chi^2 = 871.0$ ($\chi_{\nu}^2 = 2.9$), with an unphysically low 2-$\sigma$ upper limit on the mass of Kepler-50 b ($<$ 0.16 M$_{\oplus}$). 

\begin{figure}
\includegraphics[height = 6.7 in]{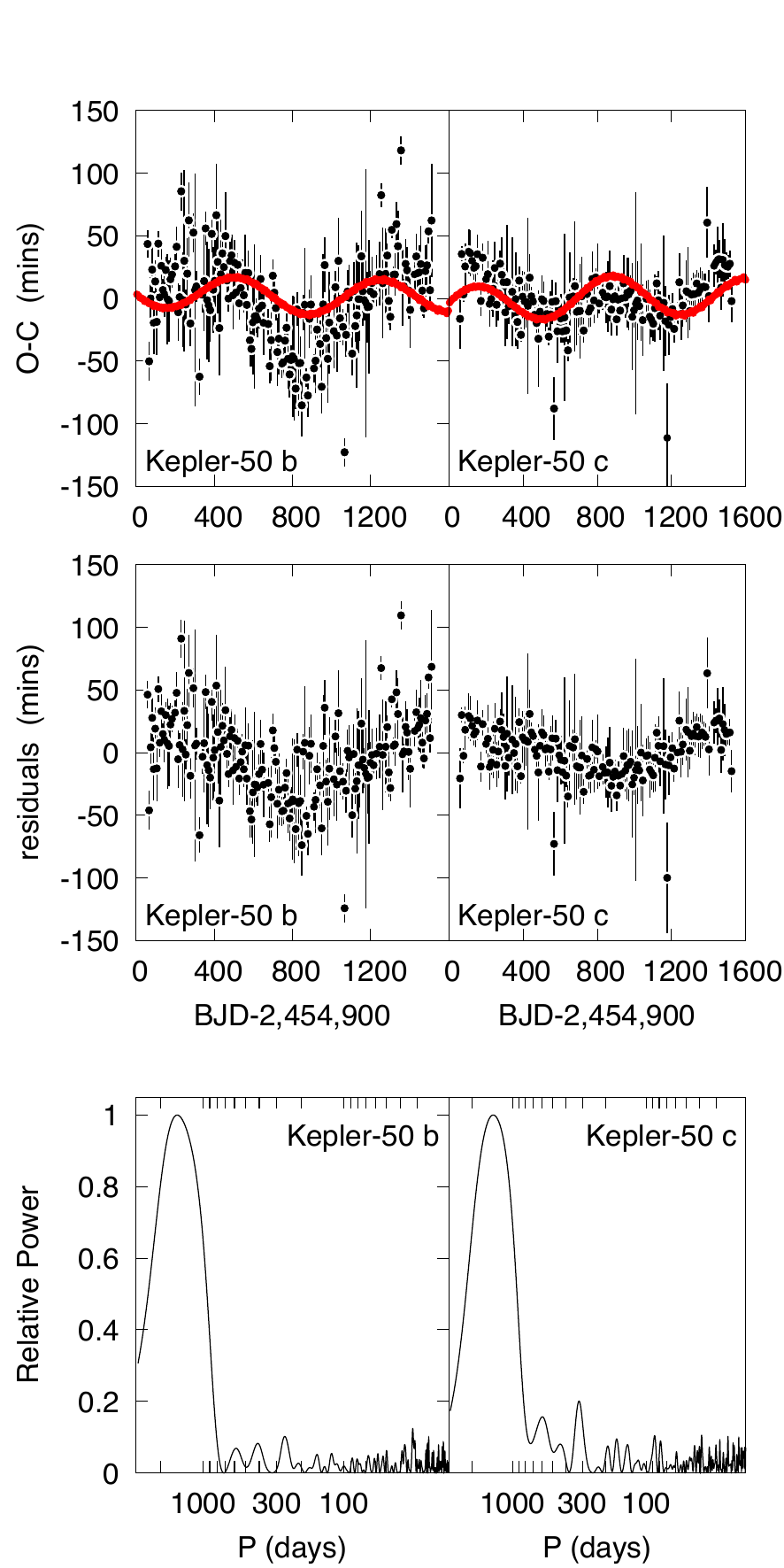}
\caption{Poor fit from a two-planet model. Top panels: Best-fit two planet models for Kepler-50 b (left) and Kepler-50 c (right) (in red points), given fixed planetary masses of 3.5 $M_{\oplus}$, over the baseline of Kepler data (with time marked below the middle panels). The black points show the data. The middle panel shows the residual TTVs. The lower panels display Lomb-Scargle periodograms of the residuals. 
}\label{Fig:2pl}
\end{figure}

Fixing the masses of both planets at 3.5 M$_{\oplus}$, we found $\chi^2 = 1157.6$ ($\chi_{\nu}^2 = 3.7$). These fits, their residuals, and Lomb-Scargle periodograms of the residuals are shown in Figure~\ref{Fig:2pl}.

Finally, we relaxed the assumption of co-planarity and allowed a free ascending node for the outer planet ($\Omega_c$). We found $\chi^2 = 1106.1$ ($\chi_{\nu}^2 = 3.7$). These poor fits motivate additional model parameters and we consider an additional planet coplanar to the transiting planets.

\section{Grid search for perturbing planet} 
\label{sec:LMfits}
The parameter space for non-transiting planets in TTV signals is, in most cases, highly degenerate, with many local minima in the goodness-of-fit. We identified regions of interest using Levenberg-Marquardt minimization (LM) with parameters initialized over a grid in orbital parameters. We refer to the putative perturbing planet as “Kepler-50x”, or “Kep50x” for brevity in some cases.
 
We initialized the fits with the best-fit 2-planet model with the masses of Kepler-50 b and c fixed at 3.5 M$_{\oplus}$. To include the best fit 2-planet model (with these masses), eccentricity vector components were limited to between $\pm 0.056$ and the mass of Kepler-50x was initialized at zero.    

We explored a range of orbital periods $P_{x}$ that included all first- and second-order resonances and excluded periods that would cause Hill Instability \citep{Gladman1993} in the system if $M_{x} = 1 M_{\oplus}$. Note that, apart from an extremely small subset of parameter space \citep{Gavino2026}, 3-planet systems need to have neighboring planets separated by much more than the two-planet Hill stability limit \citep{Lissauer2021}, so our criteria are quite inclusive even for perturbers of mass $M_{x} \ll 1 M_{\oplus}$. To resolve local minima in $P_{x}$, we used a  grid resolution of $dP_{x} = 10^{-5}P_{x}$.    

\begin{figure}
\includegraphics[width = 3.5 in]{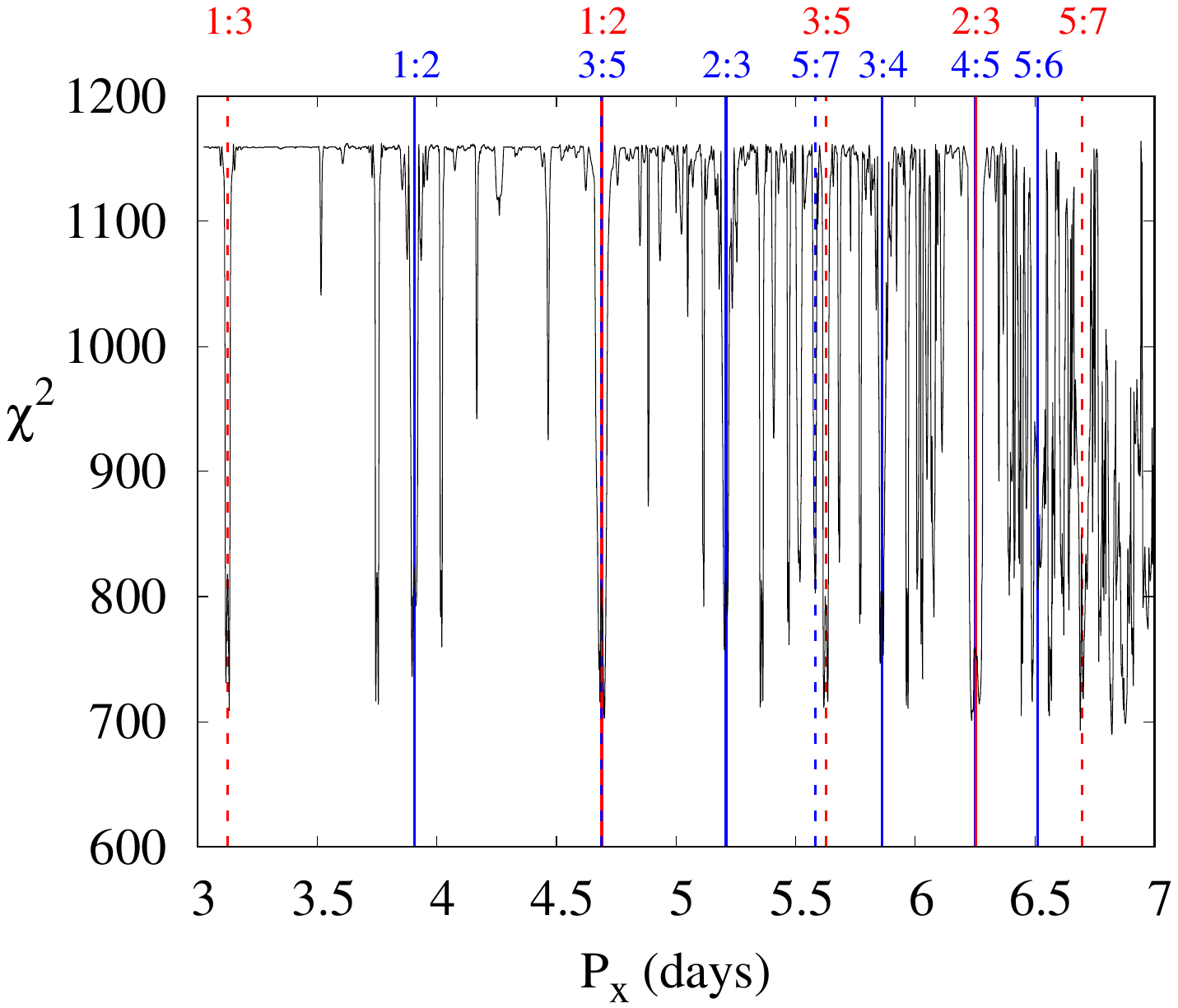}
\includegraphics[width = 3.5 in]{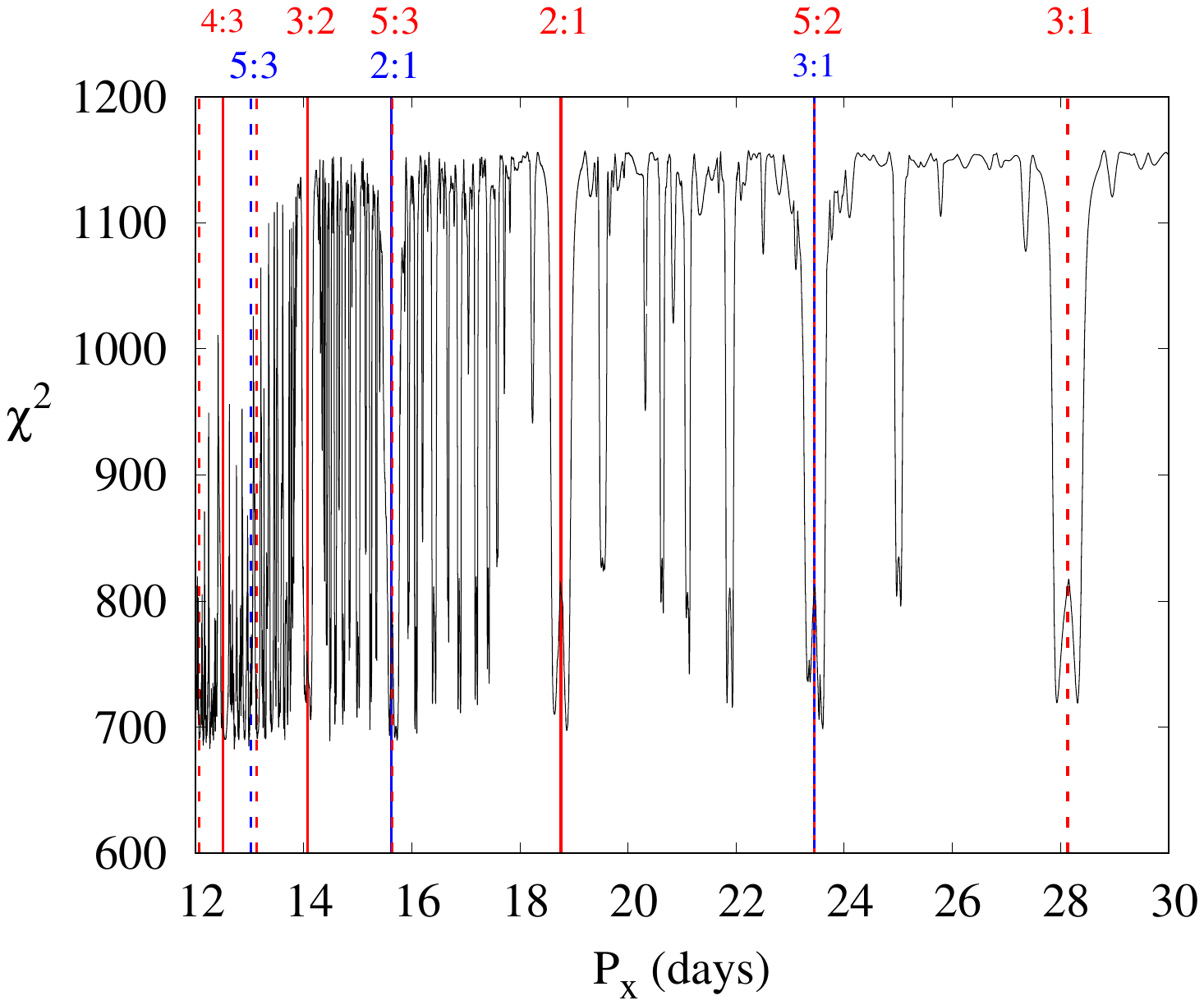}
\caption{Goodness of fit for three-planet models with the masses of the transiting planets held fixed at $3.5 M_\oplus$ over a range of fixed periods for a putative non-transiting planet Kepler-50x orbiting interior to Kepler-50 b (top) and exterior to Kepler-50 c (bottom).  The vertical lines mark first- and second-order resonances (solid and dashed respectively), with Kepler-50b (in blue) and Kepler-50c (in red).}  \label{Fig:LM}
\end{figure}

At each $P_{x}$, we tested 8 uniformly-spaced initial orbital phases, with the first phase chosen randomly, and compared the best fit of these and the parameters (other than $P_{x}$) from the best fit for the preceding $P_{x}$ in our grid in ascending order of $P_{x}$. We conducted an additional sweep of periods comparing fits to preceding $P_{x}$ in descending order along the grid. The best fit at each $P_{x}$ was recorded and the results are displayed in Figure~\ref{Fig:LM}.

There are many solutions that are isolated from one another in parameter space, and many associated with resonances with one or both transiting planets. The proximity of solutions increases as Kepler-50x is placed closer to the transiting planets. 

\begin{figure}
\includegraphics[width = 3.5 in]{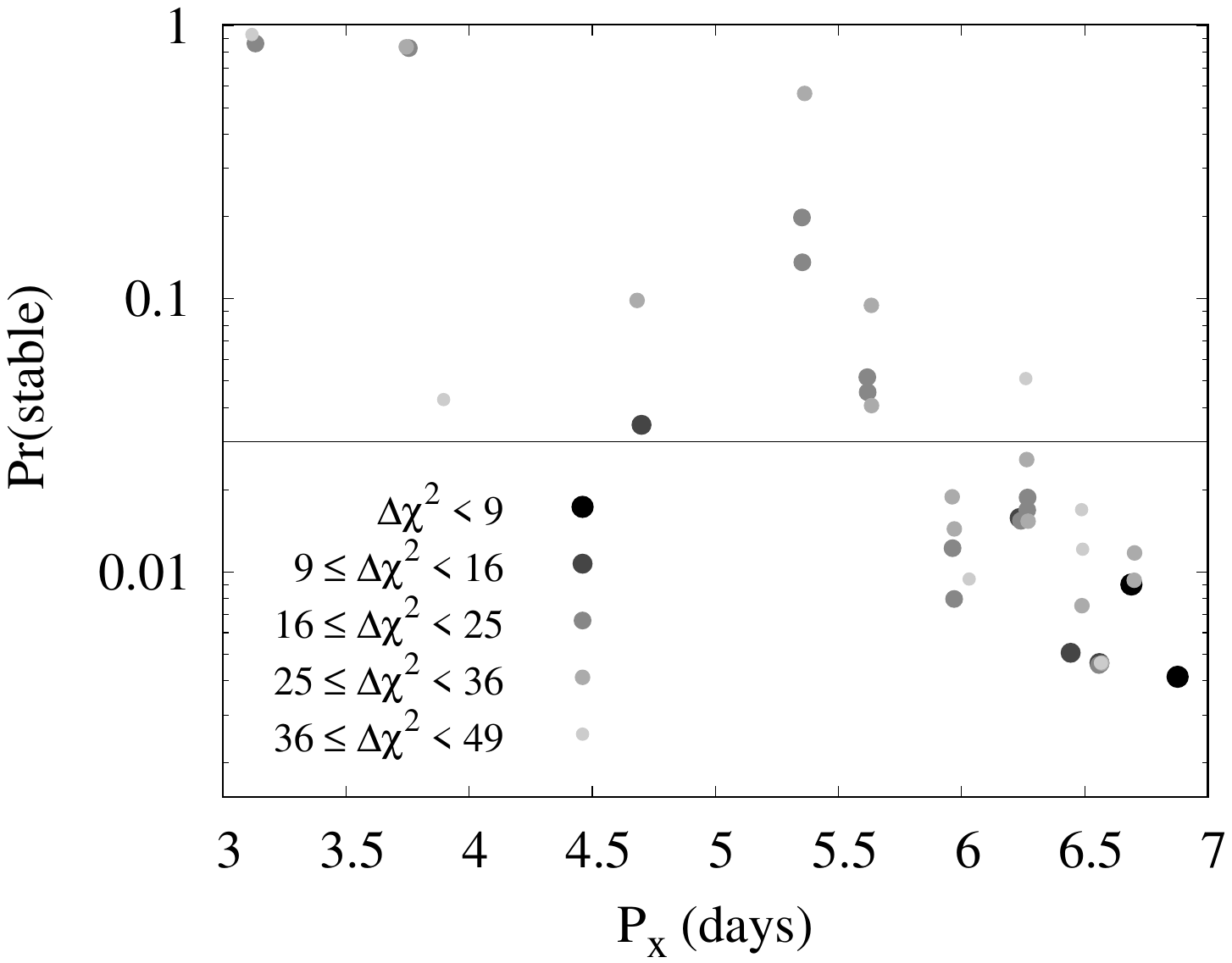}
\includegraphics[width = 3.5 in]{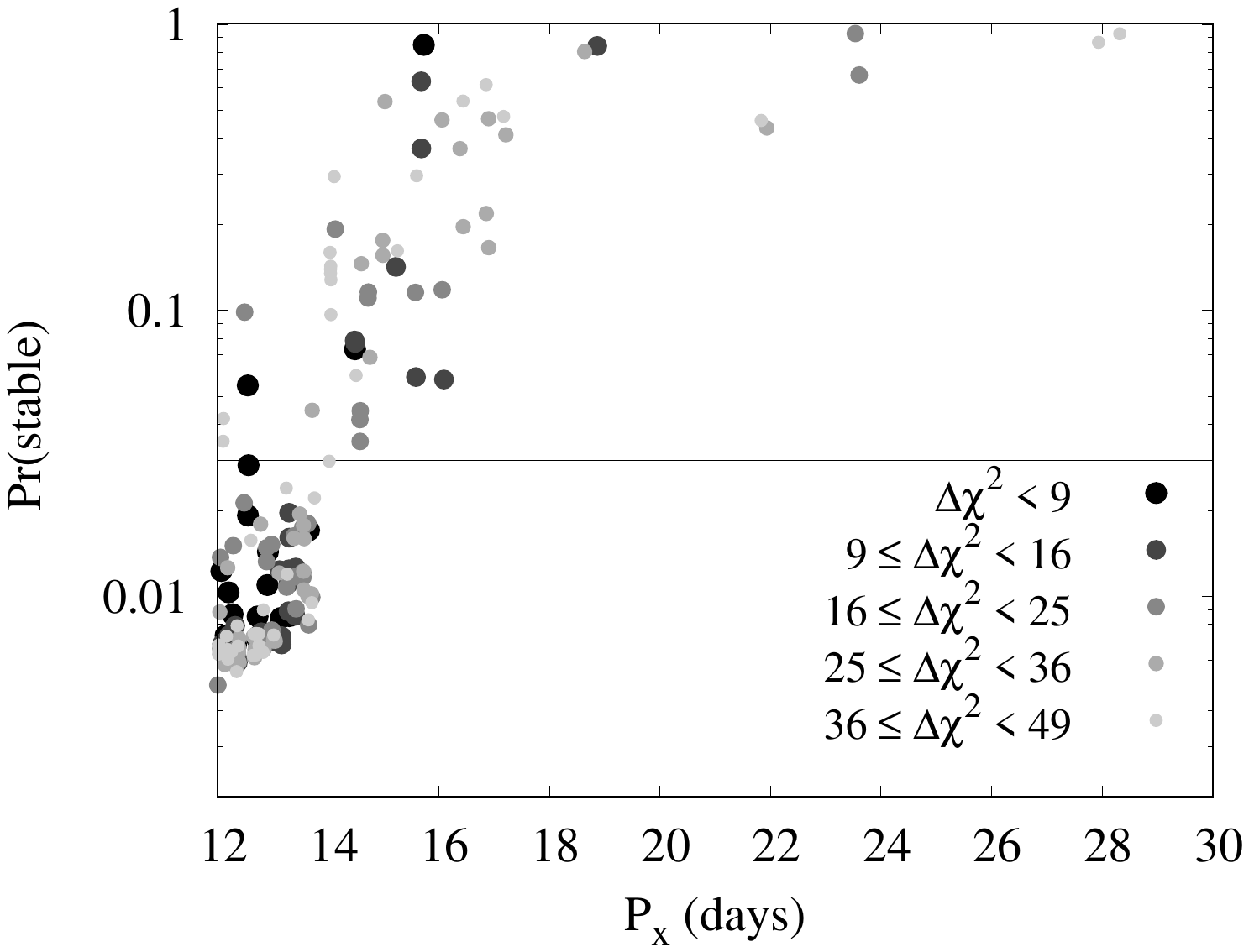}
\caption{The likelihood of long term stability of solutions according to Spock at the local minima identified in Figure~\ref{Fig:LM}. The symbol size indicates the comparative goodness-of-fit to the overall best fit, and the horizontal line marks where Pr(stable) = 0.03.}
\label{Fig:LMspock}
\end{figure}

However, as the planets are placed closer to one another in orbital period, the likelihood that the configuration is long-term stable decreases. We estimated the likelihood that the solution at each local minimum is stable for $10^9$ orbits\footnote{The planets in the Kepler-50 system have completed $\sim 10^{12}$ orbits, but tidal damping can counteract perturbations that require millions of years to destabilize the system.} by using the Stability of Planetary Orbital Configurations Klassifier (Spock) \citep{Tamayo2020} and highlight the results in Figure~\ref{Fig:LMspock}. The majority of solutions found near the overall best fit model are unlikely to be long-term stable according to Spock. We selected the regions of interest where Pr(stable) $>$ 0.03 for further analysis. 

We included all local minima where $\Delta \chi^2 < 49$ from the best fit overall. We adopt this criterion that is more conservative than a standard model comparison (using F-ratio or Bayes Information Criterion) because outlying transit times can permit a significant improvement in $\chi^2$ with a fairly small change to the TTV model (also noted by \citealt{Lammers2026}) and hence a standard cutoff in $\Delta\chi^2$ that assumes Gaussian uncertainties on measured transit times could miss otherwise viable regions of interest. Furthermore, since we used fixed masses (and bounds on eccentricity) on the transiting planets in our LM search for regions of interest, we anticipated that freeing the masses for posterior sampling at each region of interest would change the relative goodness of fit ($\Delta \chi^2$) of the best fit models at each minimum. 

\section{Posteriors} 
\label{sec:DEMCfits}
At each local minimum where Pr(stable) $>$ 0.03, we performed differential evolution MCMC, with free masses, orbital periods, orbital phases and eccentricity vectors ($h$,$k$) for all three planets. For priors, masses were positive definite, ($h$,$k$) were drawn from Gaussian distributions of width 0.02 and the first transit after epoch of the putative planet $T_{0,x}$ was bound between epoch and epoch+$P_{x}$. We launched `walkers' at the local minimum of each of the 15 identified regions of interest interior to Kepler-50 b and the 55 identified regions of interest exterior to Kepler-50 c. We took 10,000 independent samples in each region of interest. 

We integrated all samples for 100 years using Rebound \citep{Rein2012} and considered simulations where the total range in $\Phi_{bc1} = 5\lambda_b - 6\lambda_c+\varpi_{b}$ or $\Phi_{bc2} = 5\lambda_b - 6\lambda_c+\varpi_{c}$ is less than $180^{\circ}$ to be in libration. 

\begin{figure}
\includegraphics[width = 3.5 in]{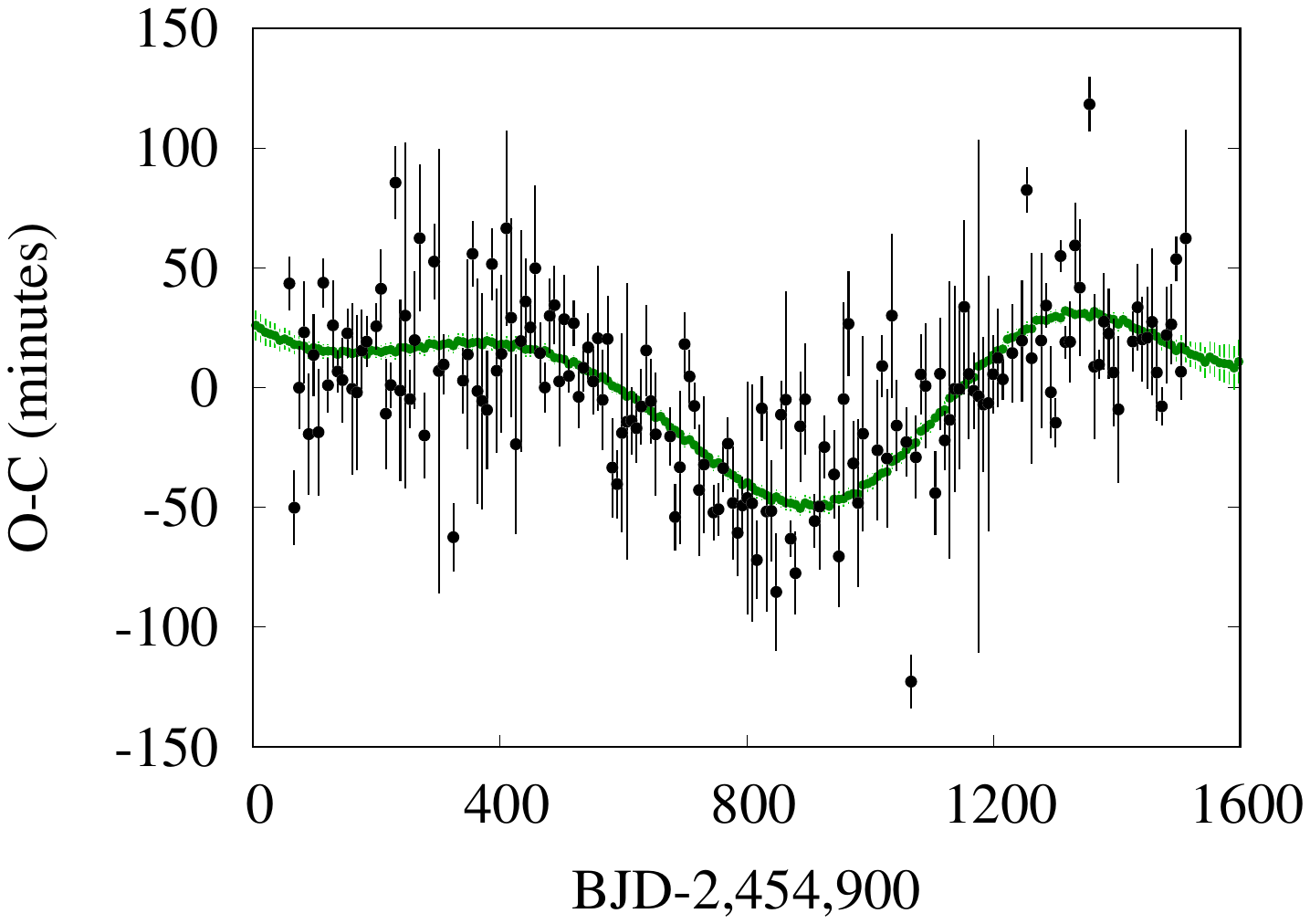}
\includegraphics[width = 3.5 in]{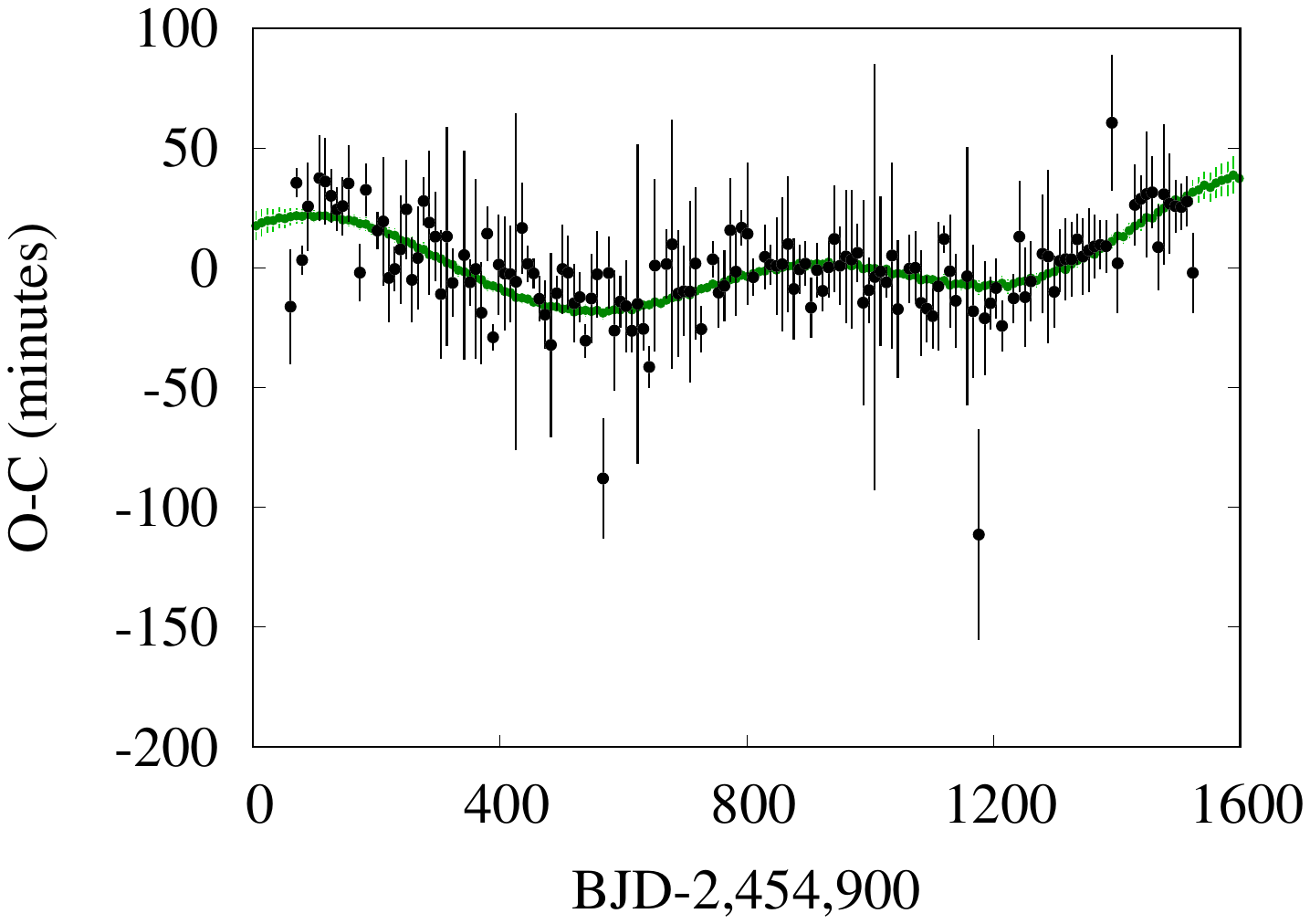}
\caption{A 3-planet TTV model. The black points are the observed transit times and their difference from a calculated linear fit. The green points with error bars are the mean and standard deviation of transit times from 1000 posterior samples. These TTV fits follow posterior sampling within the region of interest summarized in the top panel of Table~\ref{Tab:mcmc}.}
\label{Fig:3pl}
\end{figure}

For each region of interest, we noted the fraction of samples that are in libration using this criterion. We summarize the results of two regions of interest in Table~\ref{Tab:mcmc}, with the TTVs displayed in Figure~\ref{Fig:3pl}. 

In addition we estimated the probability of enduring $10^9$ orbits with Spock, and note the fraction of samples in each region of interest where the likelihood of stability is greater than 50\%. 

\begin{table*}
  \begin{center}
  \begin{tabular}{|c|c|c|c|c|c|c|c|c|}
 \hline
name & Period (d) & T$_{0}$ (d) & $e\cos \varpi$ & $e\sin \varpi$ & $\frac{M_p}{M_{\oplus}} \frac{M_{\odot}}{M_{\star}}$ &  Min($\chi^2$)  & Fr(res) & Fr(stable) \\
\hline
Kep50x & $ 4.7008^{+0.0015}_{-0.0018}$ & $  783.532^{+  0.179}_{-  0.132}$ & $ 0.012^{+0.011}_{-0.011}$ & $-0.003^{+0.009}_{-0.009}$ & $ 1.0^{+0.5}_{-0.4}$  & & & \\ 
Kep50b & $ 7.8120^{+0.0001}_{-0.0001}$ & $  785.305^{+  0.002}_{-  0.002}$ & $ 0.011^{+0.014}_{-0.013}$ & $ 0.029^{+0.011}_{-0.011}$ & $ 2.2^{+0.6}_{-0.5}$ & & & \\
Kep50c & $ 9.3771^{+0.0001}_{-0.0001}$ & $  782.573^{+  0.001}_{-  0.002}$ & $-0.022^{+0.015}_{-0.014}$ & $-0.021^{+0.011}_{-0.011}$ & $ 3.5^{+0.3}_{-0.3}$ & 682.38 & 0.9552 & 0.8747 \\
\hline  
Kep50b & $ 7.8120^{+0.0001}_{-0.0001}$ & $  785.304^{+  0.002}_{-  0.002}$ & $ 0.016^{+0.013}_{-0.014}$ & $ 0.022^{+0.013}_{-0.014}$ & $ 2.2^{+0.3}_{-0.3}$ & & & \\
Kep50c & $ 9.3771^{+0.0001}_{-0.0001}$ & $  782.573^{+  0.002}_{-  0.002}$ & $-0.013^{+0.012}_{-0.013}$ & $-0.023^{+0.012}_{-0.013}$ & $ 3.2^{+0.3}_{-0.3}$ & & & \\
Kep50x & $12.4802^{+0.0021}_{-0.0021}$ & $  785.914^{+  0.276}_{-  0.264}$ & $-0.007^{+0.012}_{-0.013}$ & $-0.003^{+0.012}_{-0.012}$ & $ 0.6^{+0.6}_{-0.4}$ & 690.72 & 0.8543 & 0.1014 \\
\hline
\end{tabular}
\caption{Summary statistics for model parameters in two examples of the regions of interest that are summarized in Figures~\ref{Fig:summaryplots_xbc} and \ref{Fig:summaryplots_bcx}. The table also compares the minimum $\chi^2$ of samples between the two regions, the fraction of samples that are found to be in 2-body libration between Kepler-50 b and c, and the fraction of samples that are stable according to Spock.} \label{Tab:mcmc}
\end{center}
\end{table*}

\begin{figure}
  \centering
\includegraphics[width=0.48 \textwidth]{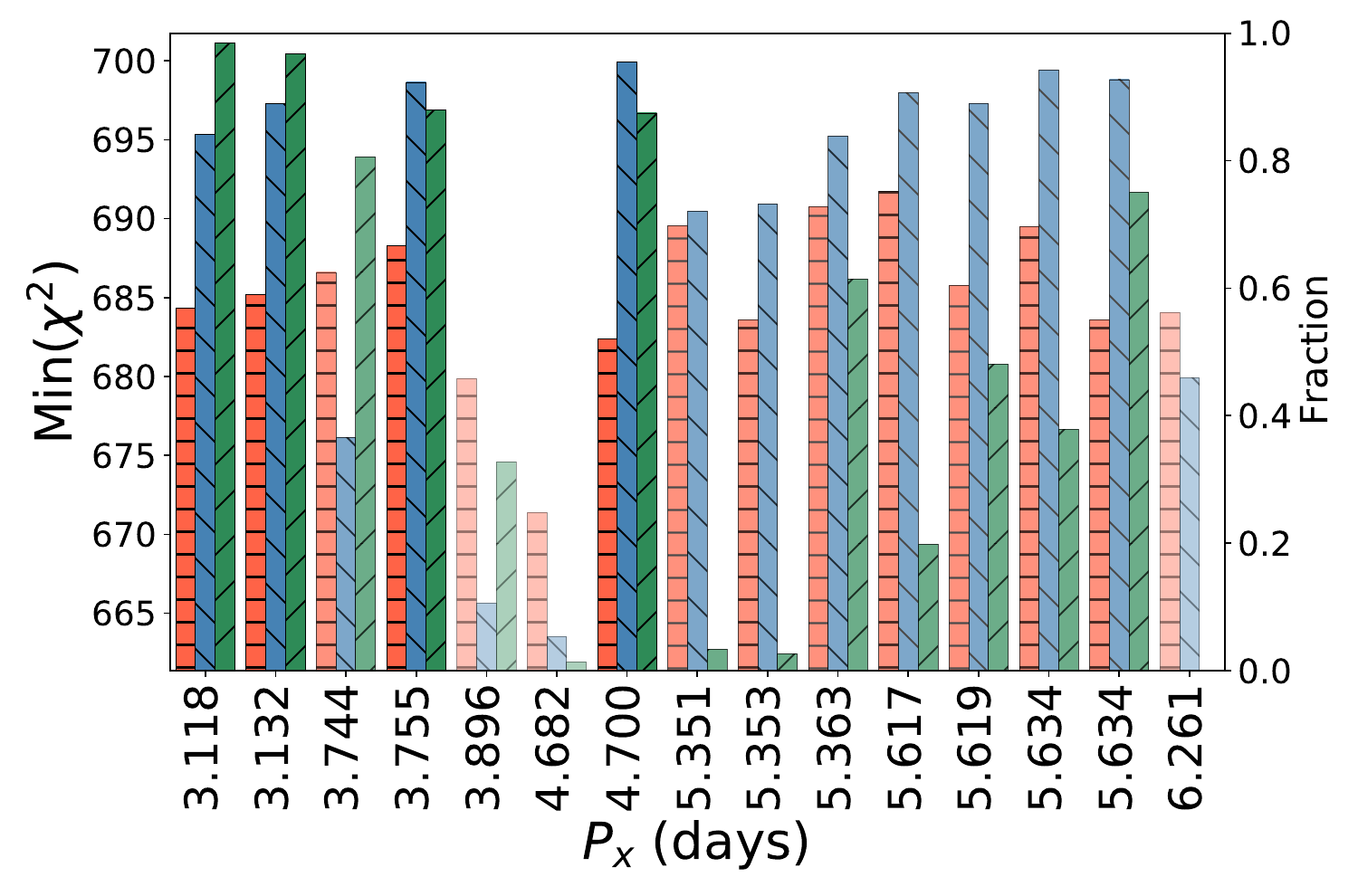}
  \caption{Summary of regions of interest interior of Kepler-50 b: Min($\chi^2$) (red), resonant fraction (Fr(res) in blue), and stable fraction (Fr(stable) in green). Highlighted are the regions where Fr(res) $>0.8$ and Fr(stable) $>$ 0.8, where a model is considered stable if Spock estimated the likelihood of surviving 10$^{9}$ orbits to be $>$ 0.5.}
  \label{Fig:summaryplots_xbc}
\end{figure}

\begin{figure*}
\includegraphics[width=0.99 \textwidth]{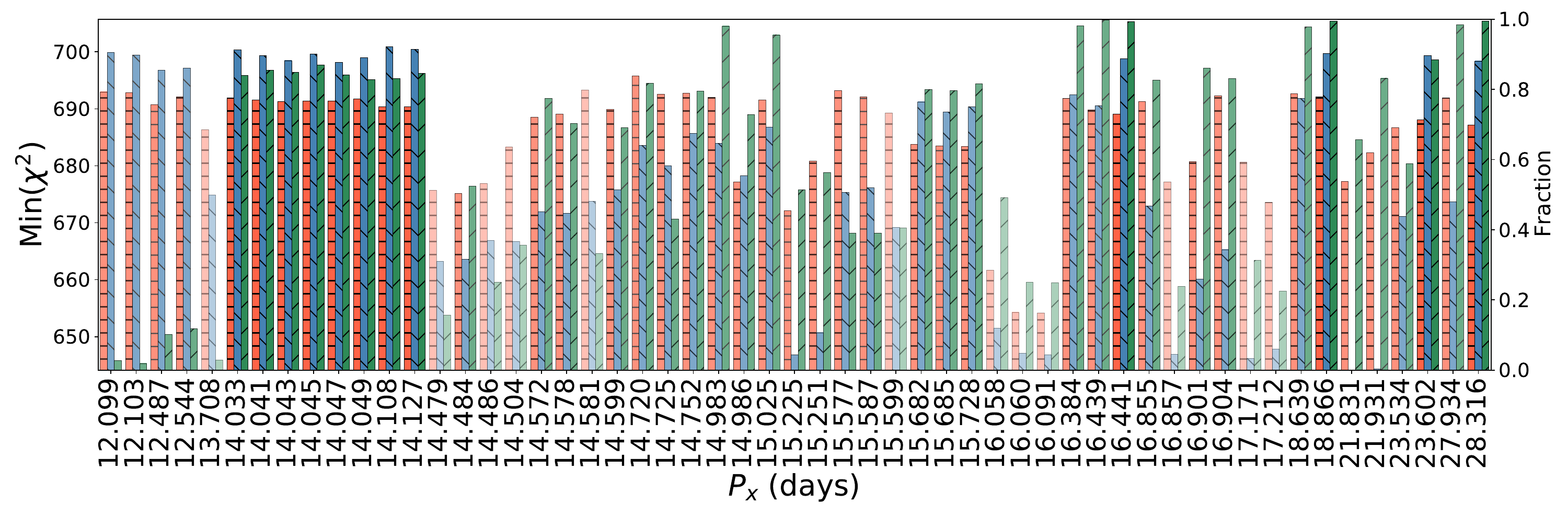}
\caption{Summary of regions of interest exterior of Kepler-50 c. Min($\chi^2$) (red), resonant fraction (Fr(res) in blue), and stable fraction (Fr(stable) in green). Highlighted are the regions where Fr(res) $>0.8$ and Fr(stable) $>$ 0.8, where a model is considered stable if Spock estimated the likelihood of surviving 10$^{9}$ orbits to be $>$ 0.5.}
\label{Fig:summaryplots_bcx}
\end{figure*}

We summarize the results of each region of interest interior to Kepler-50 b (Figure~\ref{Fig:summaryplots_xbc}) and exterior to Kepler-50 c (Figure~\ref{Fig:summaryplots_bcx}), in terms of the minimum $\chi^2$, the fraction of samples that are in 2-body libration, and the fraction of samples that are stable.

With 312 data points and 15 free parameters, our best-fit model gives $\chi^2 = 654.1$ and a reduced $\chi^2_{\nu} = 2.2$. This is higher than a statistically ideal fit to the transit timing data, and is likely due to the few extreme outliers that can be seen in Figure~\ref{Fig:3pl}. We note that there are 7 outliers beyond 4-$\sigma$, that add $\approx$261 to the total $\chi^2$ value. 

There is no clear preference for a better fit with samples in libration or likely to be stable. Furthermore, while the fraction of samples that are stable is clearly less for regions where Kepler-50 x is placed near the transiting planets (as we found with the LM fits), the remaining regions of interest show that unstable samples are well-mixed overall.


\subsection{Stability weakly constrains posteriors}
\begin{figure}
\includegraphics[width = 2.0 in]{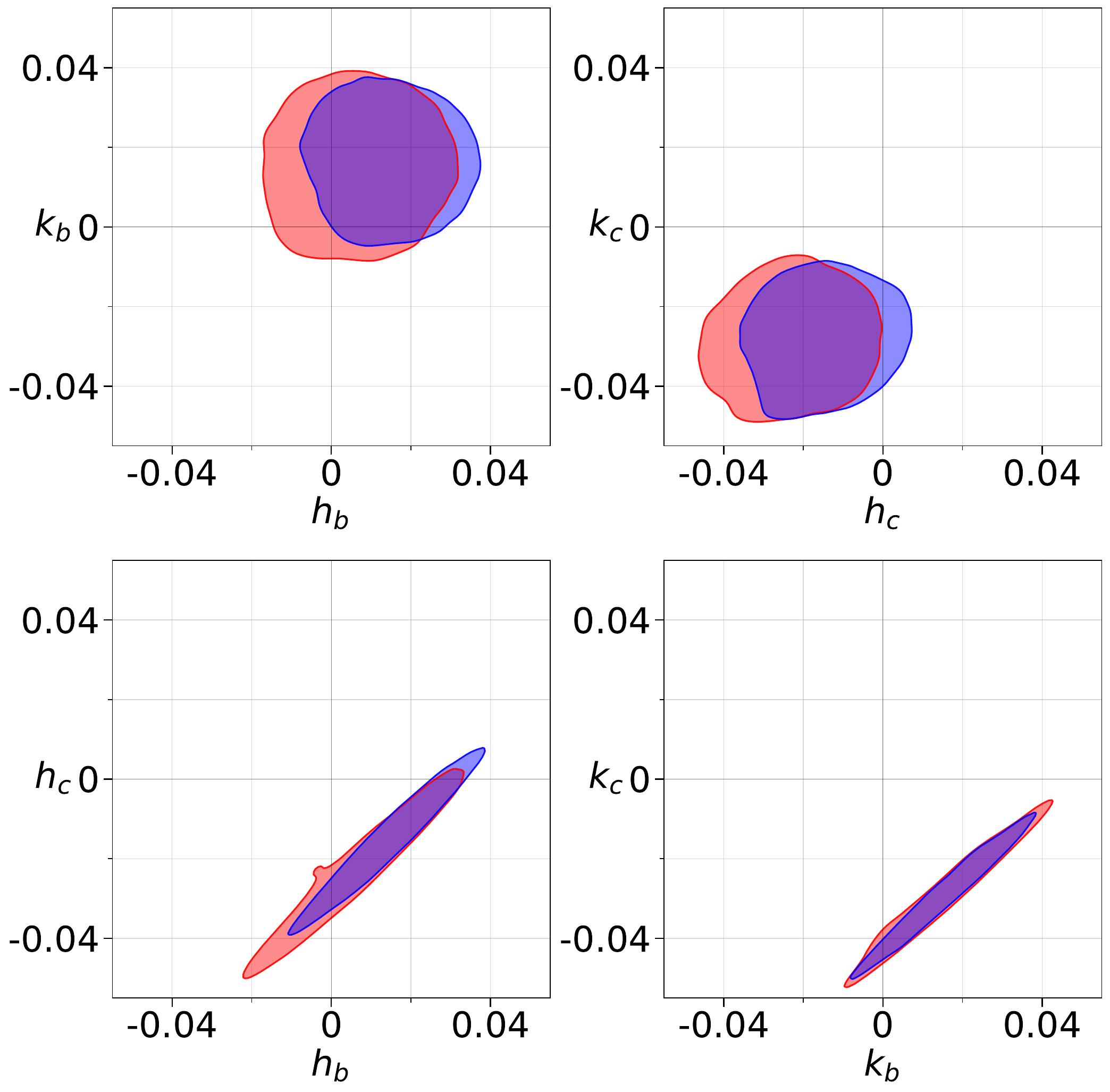}
\includegraphics[width = 1.3 in]{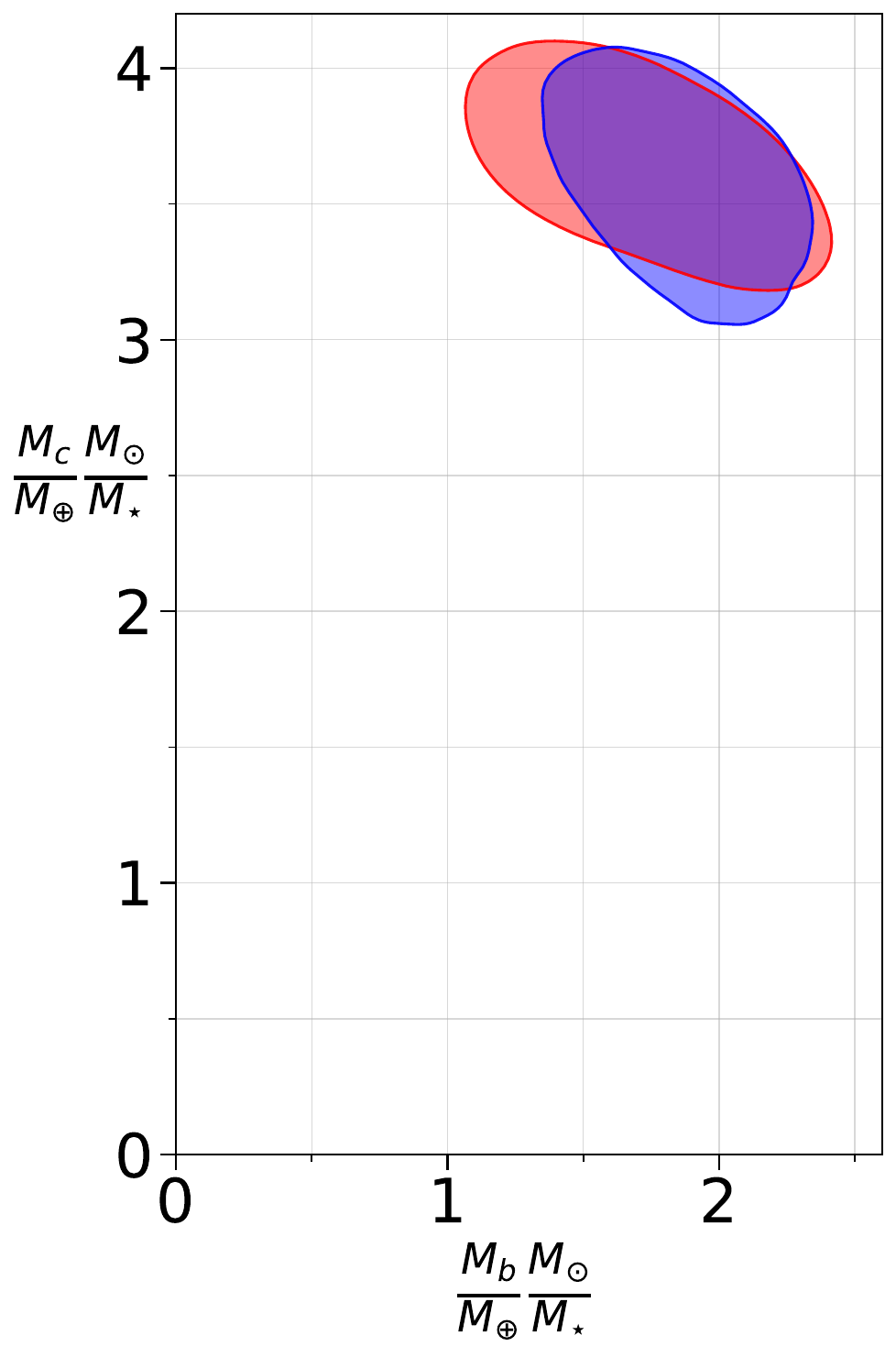}
\caption{Joint posteriors for the most stable and least stable samples. Filled contours enclose the 68.3\% (1$\sigma$) KDE  credible region of posterior samples in stability categories. In red Pr(stable) is less than 20\%, while in blue  Pr(stable) $>$ 80\% according to Spock.}
\label{Fig:Stability}
\end{figure}

We sorted the 700,000 samples by their likelihood of stability according to Spock. 417,221 ($\approx 60\%$) were likely stable. Figure~\ref{Fig:Stability} shows 2D Kernel Density Estimators (KDEs) to illustrate joint posteriors in dynamical masses and eccentricity vector components for Kepler-50 b and c. We compare the joint posteriors of samples where Spock estimated a likelihood of stability less than 20\%, against those where Spock estimated a likelihood of stability above 80\%. The close agreement between the joint posteriors highlights how well-mixed stable and unstable solutions are, as noted by \citet{Lammers2026} for the systems considered in that study (which did not include Kepler-50). 

\subsection{Samples in libration}
Of our 417,221 posterior samples with a likelihood of long-term stability $> 0.5$ according to Spock, 285,138 were found to be in 2-body libration between Kepler-50 b and c. We considered a sample to be in 2-body libration if either of the resonant arguments: $\Phi_{bc1} = 5\lambda_b - 6\lambda_c + \varpi_b$, or $\Phi_{bc2} = 5\lambda_b - 6\lambda_c + \varpi_c$ had a total range $< 180^{\circ}$ over a 100 year simulation using Rebound \citep{Rein2012}. We adopt this subsample ($\approx 68\%$ of the stable samples) as our nominal result in this study. 

\begin{figure}[h]
  \centering
  \includegraphics[width = 3.3 in]{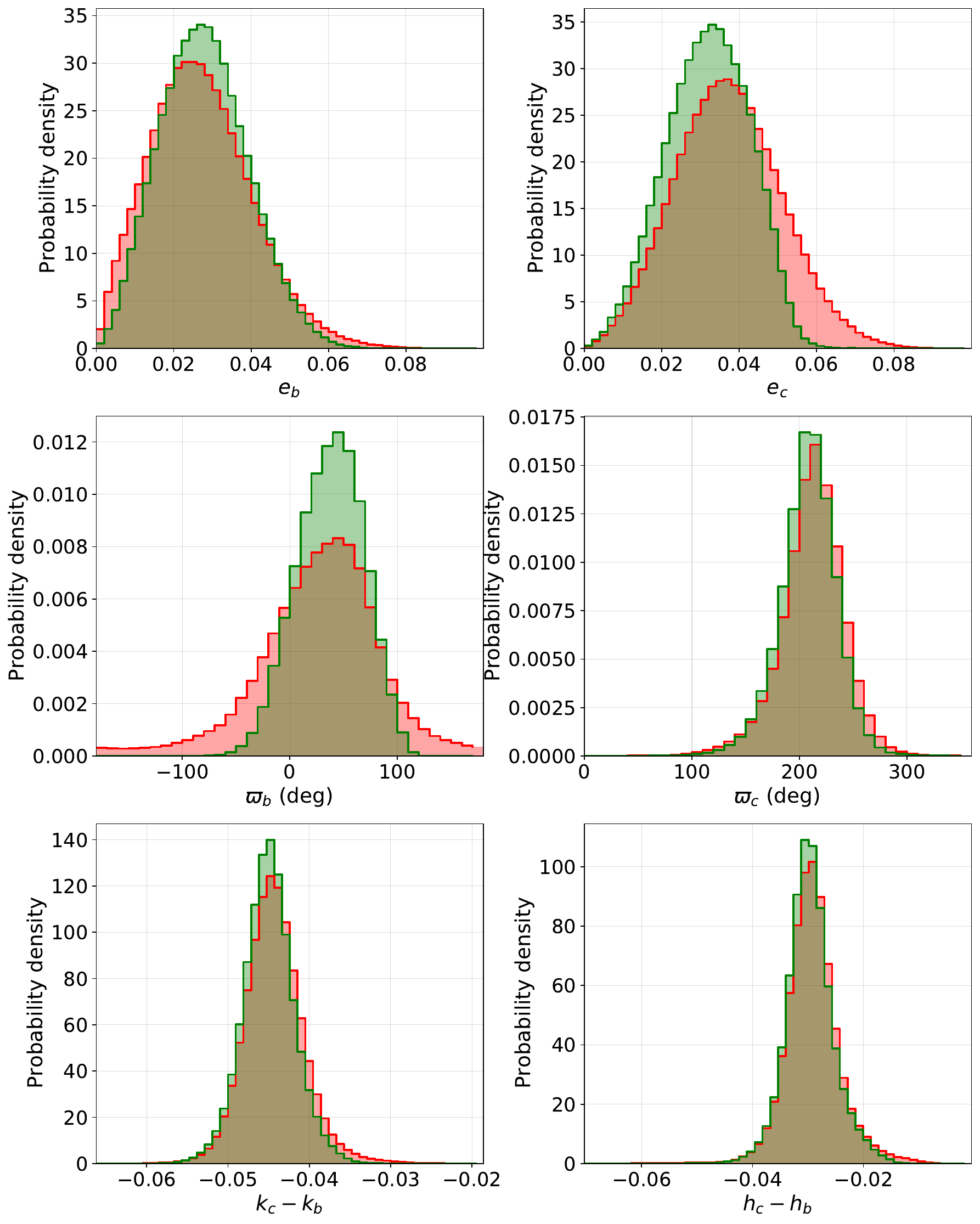}
  \caption{Area-normalized posterior histograms of orbital eccentricity ($e$), argument of pericenter ($\varpi$) for Kepler-50 b and c, and the difference in eccentricity vector component $h_c-h_b$ and $k_c-k_b$  for two sample selections:
  all posterior samples (red, 700,000 samples);
  samples with $\mathrm{Pr(stable)}>0.5$ that are also in the $bc$ 2-body
  resonance (green, 285,138 samples).}
  \label{fig:histo_ecc}
\end{figure}

The fully marginalized eccentricity posteriors in Figure~\ref{fig:histo_ecc} show tight upper bounds on the individual eccentricities of the transiting planets, although this is partially driven by our prior. Nevertheless, the stable samples have a moderately lower upper bound on eccentricity, particularly for Kepler-50 c.

More importantly, we find that relative eccentricity is strongly detected for the pair of transiting planets (see Table~\ref{Tab:mcmc}), and that their orbit pericenters are likely anti-aligned. Some samples even have crossing orbits but remain stable due to the 6:5 commensurability.

From the samples in 2-body libration, we sought samples in 3-body libration, where, in addition to the 2-body libration state identified between Kepler-50 b and c, we also find a 2-body libration state between either x and b for the search interior to Kepler-50 b, or between c and x for the search exterior to Kepler-50 c, and the 3-body resonant argument $\Phi_{bcx} = p \lambda_b -(p+q) \lambda_c + q \lambda_{x}$, where $p$ and $q$ are small integers, is in libration. We searched for first- and second-order resonances between Kepler-50 x and b, and Kepler-50 c and x, and again included the condition that Pr(stable) $> 0.5$. We found 208 samples in 3-body libration with a first-order resonance where Kepler-50 x is exterior to c. These are mostly in the 4:3 resonance between Kepler-50 c and x (163 samples), but also include samples in 3:2 (38 samples) and 2:1 (7 samples). The posteriors for these are in Table~\ref{Tab:samples} labeled Case A. We also identified 87 samples in 3-body libration with a second-order resonance between Kepler-50 c and x (Case B). All of these were in the 9:7 resonance between Kepler-50 c and x such that the resonant argument in libration $\Phi_{bcx} = 10\lambda_b - 19\lambda_c+9 \lambda_x$. We illustrate examples of Case A and Case B in 3-body libration in the Appendix in Figure~\ref{Fig:Phi_t}.  

We found no 3-body chains that met our criteria with Kepler-50 x interior to Kepler-50 b. The resonant arguments and the number of samples and their best-fit $\chi^2$-values are listed in Table~\ref{Tab:3bodyResArgs}

\begin{figure}
\includegraphics[width = 2.0 in]{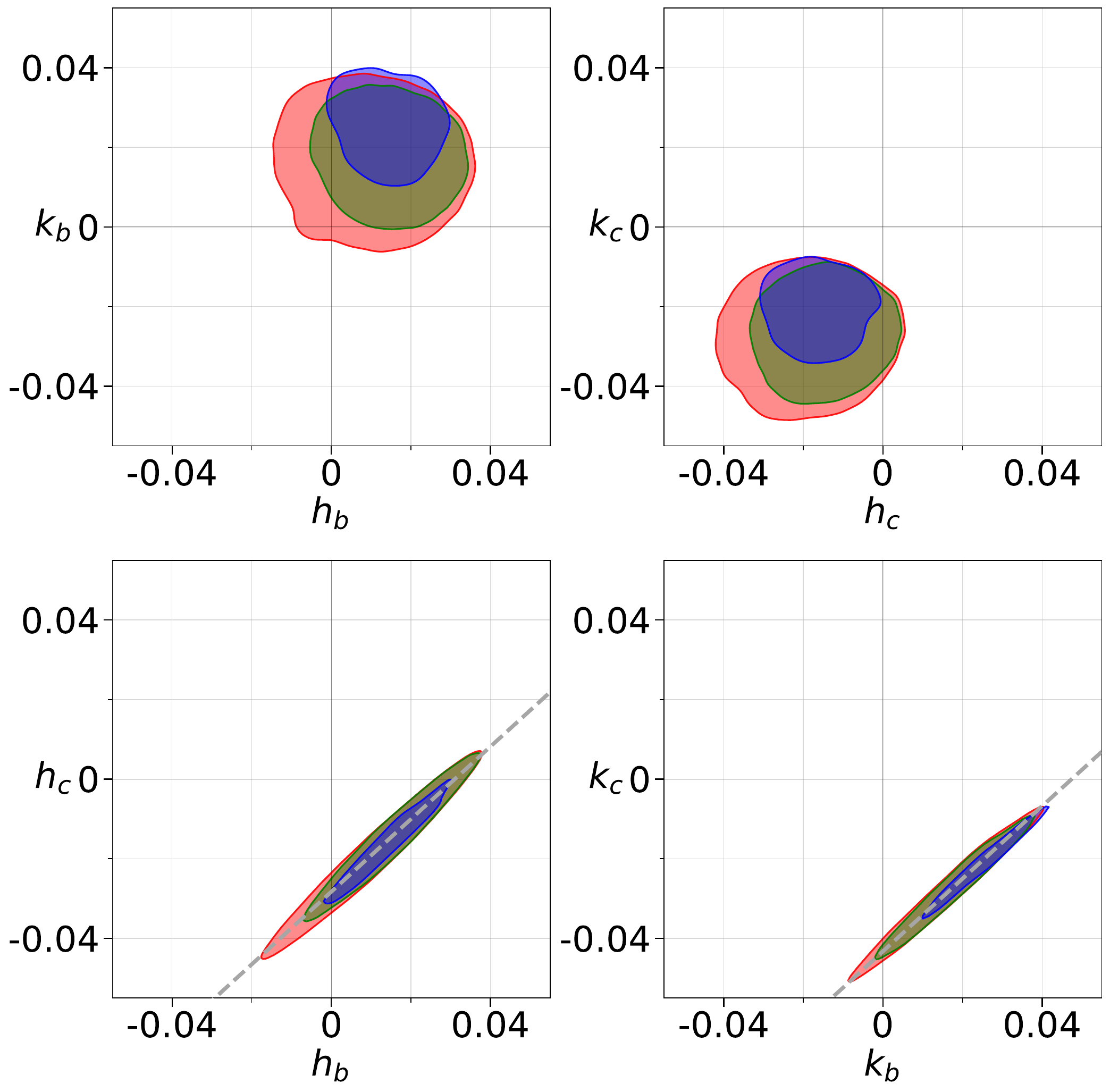}
\includegraphics[width = 1.3 in]{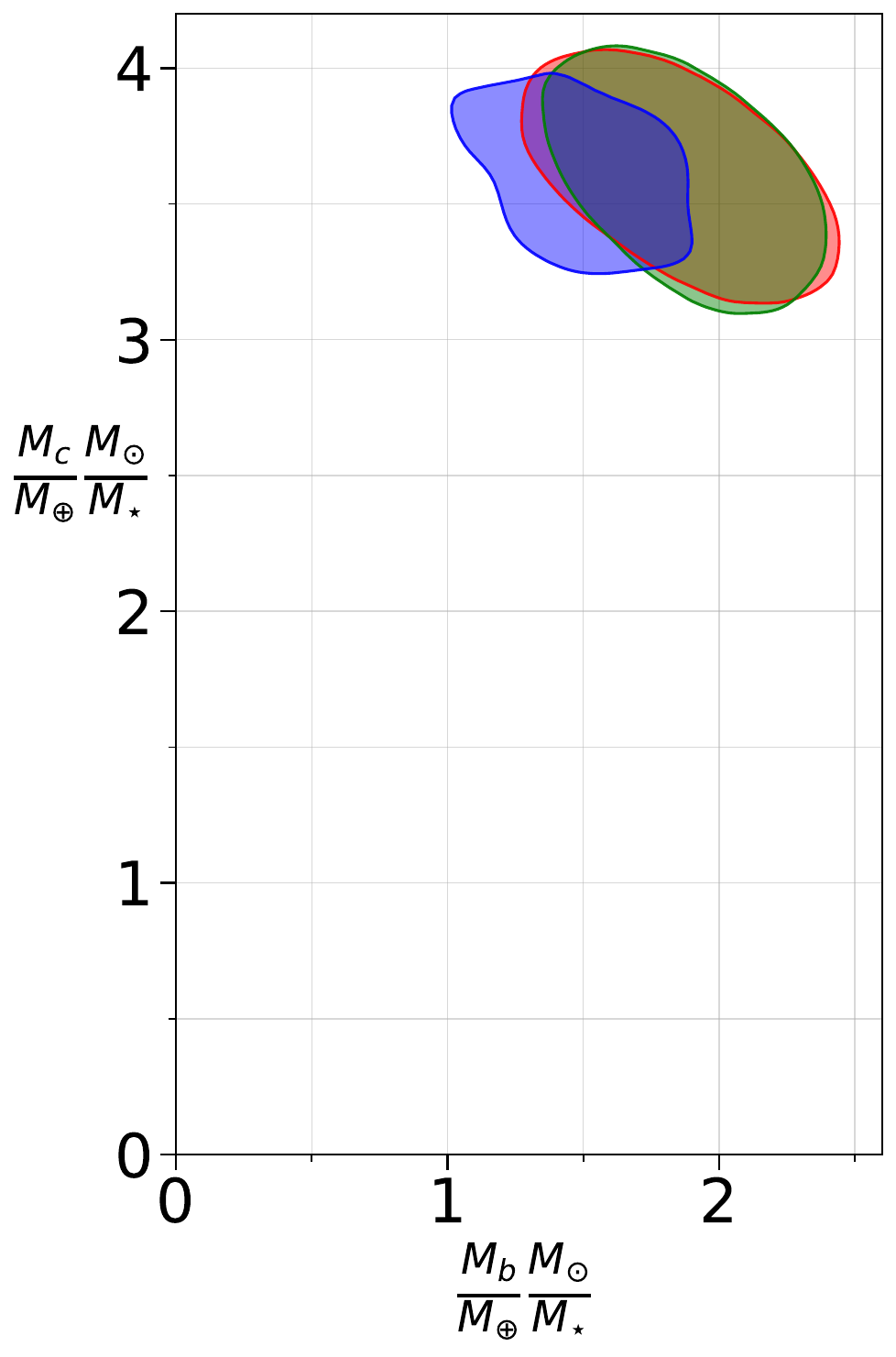}
\caption{1$\sigma$ joint-posterior contours of eccentricity vector components $h=e\sin\varpi$ and $k=e\cos\varpi$ and dynamical masses  $(M_b/M_\oplus)(M_\odot/M_\star)$ and $(M_c/M_\oplus)(M_\odot/M_\star)$ for Kepler-50b and Kepler-50c, from three nested sample selections: all posterior samples (red, 700,000 samples); samples with $\mathrm{Pr(stable)}>0.5$ that are also in the $bc$ 2-body resonance (green, 285,138 samples); and samples librating in the 3-body resonance that are also in both the $bc$ and 1st-order $cx$ 2-body resonances (blue, 208 samples). The dashed grey line indicates the expected slope $\frac{dh_{c}}{dh_b} = \frac{dk_{c}}{dk_b}$, centered on our posteriors.}
\label{Fig:resKDEs1}
\end{figure}

\begin{figure}
\includegraphics[width = 2.0 in]{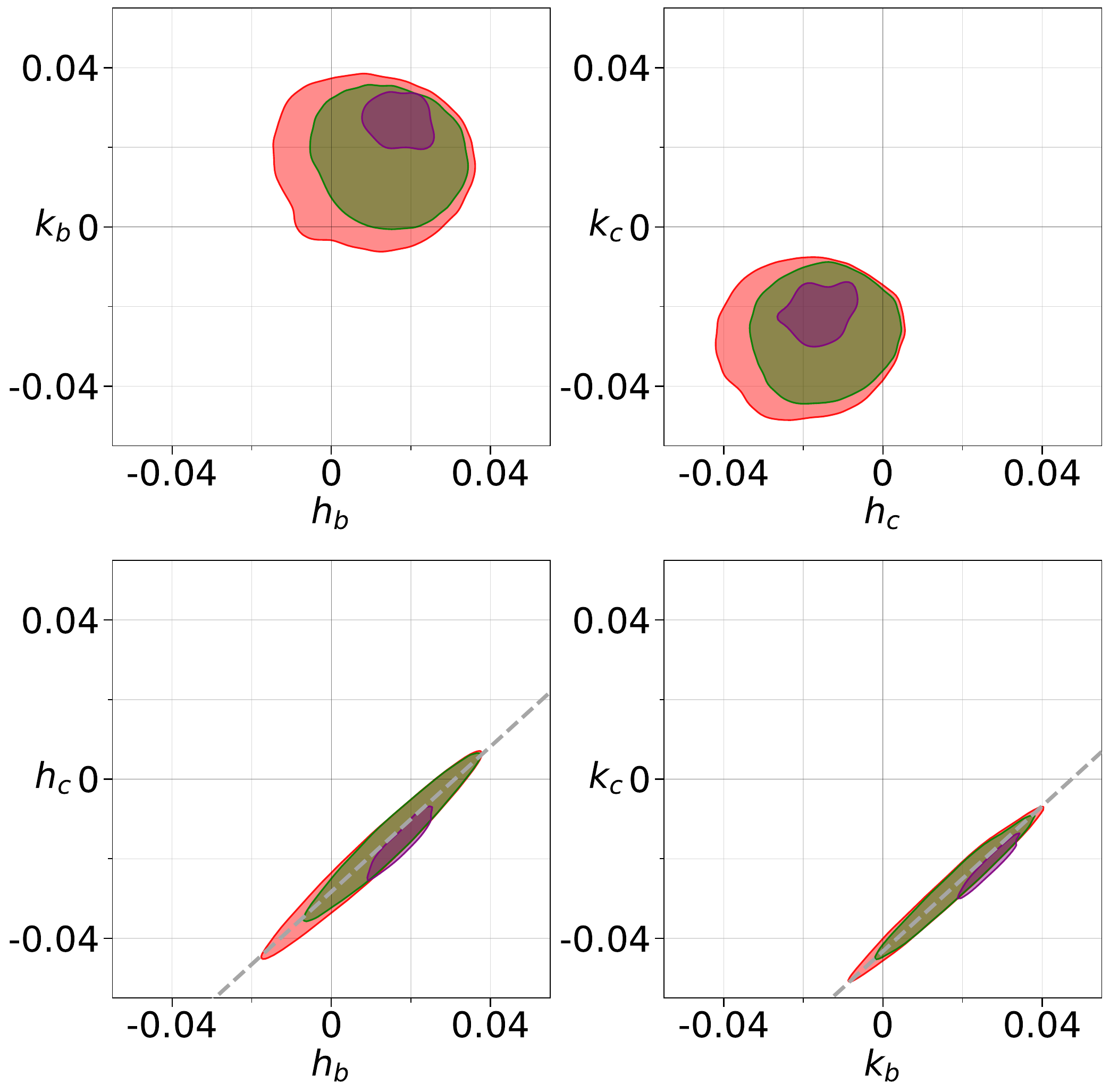}
\includegraphics[width = 1.3 in]{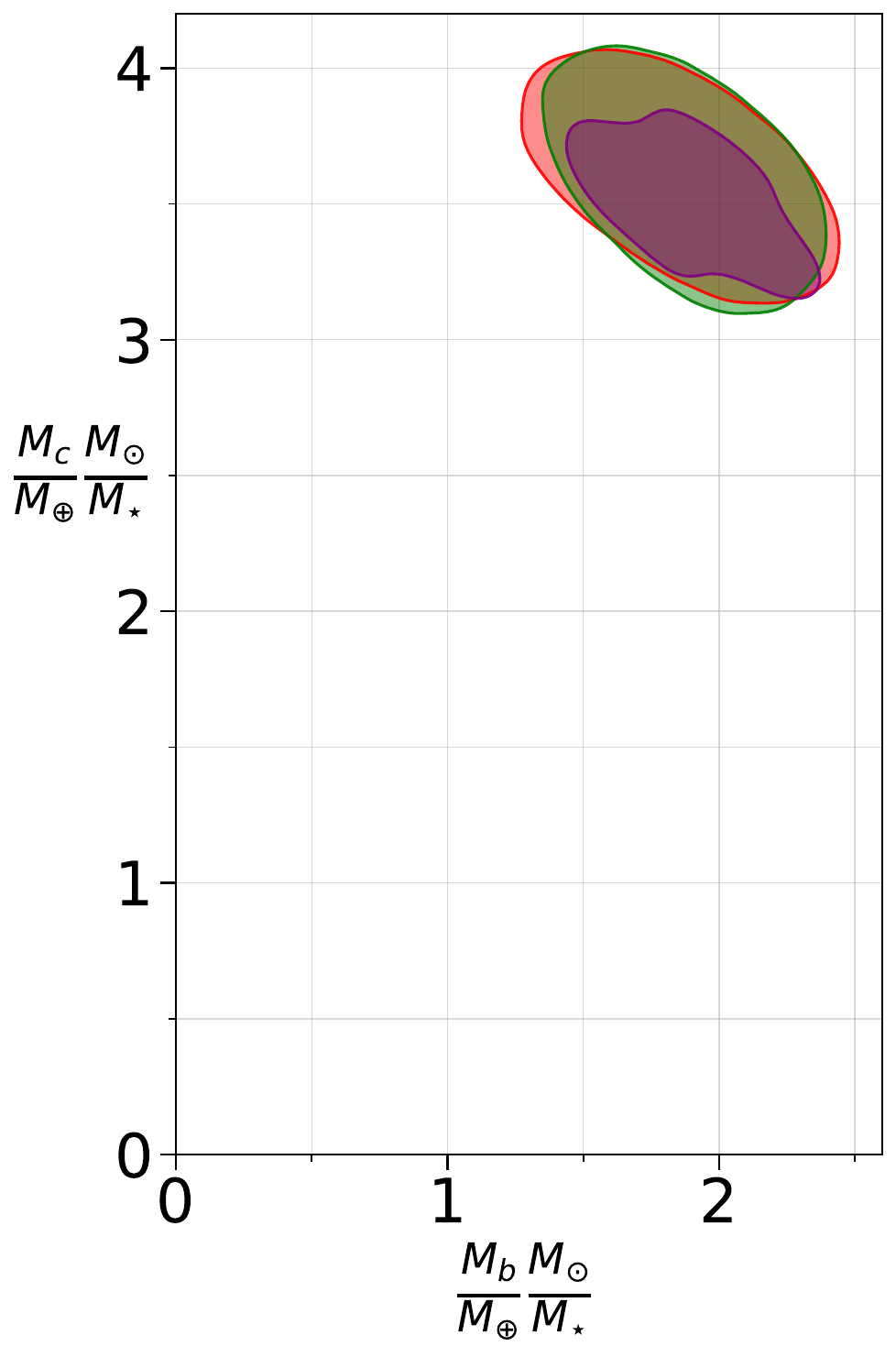}
\caption{1$\sigma$ joint-posterior contours of eccentricity vector  components $h=e\sin\varpi$ and $k=e\cos\varpi$ and dynamical masses $(M_b/M_\oplus)(M_\odot/M_\star)$ and $(M_c/M_\oplus)(M_\odot/M_\star)$ for Kepler-50b and Kepler-50c, from three nested sample selections: all posterior samples (red, 700,000 samples); samples with $\mathrm{Pr(stable)}>0.5$ that are also in the $bc$ 2-body resonance (green, 285,138 samples);  and samples librating in the 3-body resonance that are also in both the $bc$ and 2nd order $cx$ 2-body resonances (purple, 87 samples).}
\label{Fig:resKDEs2}
\end{figure}

Figures~\ref{Fig:resKDEs1} and~\ref{Fig:resKDEs2} show 2D KDEs to illustrate joint posteriors in dynamical masses and eccentricity vector components for Kepler-50 b and c, for different subsets of our posterior samples, by detections of 2-body and 3-body resonances, where Figure~\ref{Fig:resKDEs1} includes first-order resonances with Kepler-50x and Figure~\ref{Fig:resKDEs2} includes second-order resonances with Kepler-50x.

The samples in 2-body libration have slightly narrower posteriors in the masses and eccentricity vector components of b and c. However, samples in 2-body libration that are also in 3-body libration are more tightly constrained in mass and eccentricity vector components (see Table~\ref{Tab:samples}). 

Nevertheless, since there are so few sub-samples in 3-body libration (and there is a higher minimum $\chi^2$ in each sub-sample where we found 3-body libration), we report the posteriors here but adopt the samples in 2-body libration only as our main result. 

In addition, the joint posteriors in Figures \ref{Fig:resKDEs1} and \ref{Fig:resKDEs2} better illustrate the strong detection of non-zero relative eccentricity. If either of the planets is on a circular orbit, the other likely has eccentricity $\gtrsim 0.05$. The joint posteriors between $h_b = e_b\sin\varpi_b$ and $h_c = e_c\sin\varpi_c$ (as well as between $k_b = e_b\cos\varpi_b$ and $k_c = e_c\cos\varpi_c$) show that non-zero relative eccentricity is strongly detected, since if $h_b = 0$, $h_c \approx -0.03 $ or if $h_c = 0$, $h_b \approx -0.03$. Similarly, if $k_b = 0$, $k_c \approx -0.04 $ or if $k_c = 0$, $k_b \approx 0.05$. 

Hence, there is a tight constraint in the joint posteriors from the correlation between $h_{b}$ and $h_{c}$, as well as between $k_{b}$ and $k_{c}$, which is missed from the simple posterior of relative-eccentricity shown in Figure~\ref{fig:histo_ecc}. The correlation is expected from the complex sum of free eccentricities to which TTVs are sensitive. \citet{Lithwick2012} derived coefficients of the disturbing function that determine the complex sum of free eccentricities to leading order in terms of the distance from resonance $\Delta$, where $\Delta = \frac{P'}{P} \frac{j-1}{j} -1$. \citet{Lithwick2012} calculated the coefficients for $j:j-1$ near-resonances up to $j = 5$. Following their procedure, we find that for the 6:5 resonance, the coefficients are $-f =  4.456 -  30.063 \Delta$ and $g =  4.885 -  29.751 \Delta$. Hence for Kepler-50 b and c, with $\Delta =   0.000287$, the slope $\frac{dh_{c}}{dh_b} = \frac{dk_{c}}{dk_b} = 0.912$. This is shown in Figures~\ref{Fig:resKDEs1} and \ref{Fig:resKDEs2} as the dashed slope that traces the 2D KDEs precisely. 

\begin{table*}
  \begin{center}
  \begin{tabular}{|c|c|c|c|c|c|c|c|}  
\hline
Planet & Period (d) & $T_0$  & $k = e\cos\varpi$ & $h = e\sin\varpi$ & $k_c - k_b$ & $h_c - h_b$ & $\frac{M_p}{M_\oplus}\frac{M_\odot}{M_\star}$\\
\hline
\multicolumn{8}{|c|}{All samples (700,000)   $\min\chi^2$ = $654.133$ } \\
\hline
Kep-50\,b & $ 7.8120 \pm 0.0003$ & $ 785.304 \pm 0.002$ & $ 0.017 \pm 0.015 $ & $ 0.010 \pm 0.017 $ & $-0.044 \pm  0.003$ & $-0.029 \pm 0.004 $ & $ 1.9 \pm 0.4$  \\
Kep-50\,c & $ 9.3772 ^{+0.0007}_{-0.0005}$ & $ 782.573 \pm 0.002 $ & $-0.028 \pm 0.014 $ & $-0.019 \pm 0.016$ &  &  & $ 3.6^{+0.3}_{-0.4}$  \\
\hline
\multicolumn{8}{|c|}{Nominal result. 6:5 librating \& Pr(stable)\,$>$\,0.5 (285,138 samples)  $\min\chi^2$ = $655.014$ }  \\
\hline
Kep-50\,b & $ 7.8121 \pm 0.0002 $ & $ 785.304 \pm 0.002 $ & $ 0.019 \pm 0.012$ & $ 0.015^{+0.014}_{-0.013}$ & $-0.045 \pm 0.003 $ & $-0.030 \pm 0.004$ & $ 1.9^{+0.4}_{-0.3}$ \\
Kep-50\,c & $ 9.3772 \pm 0.0004 $ & $ 782.573 \pm 0.002 $ & $-0.026^{+0.012}_{-0.011}$ & $-0.015^{+0.013}_{-0.012}$ &  &  & $ 3.6^{+0.3}_{-0.4}$  \\
\hline
\multicolumn{8}{|c|}{3-body Case~A (208 samples) $\min\chi^2$ = $696.435$ } \\
\hline
Kep-50\,b & $ 7.8121 \pm 0.0001$ & $ 785.304 \pm 0.002 $ & $ 0.026^{+0.009}_{-0.010}$ & $ 0.015^{+0.009}_{-0.010}$ & $-0.046 \pm 0.002$ & $-0.030^{+0.002}_{-0.003}$ & $ 1.5^{+0.2}_{-0.3}$  \\
Kep-50\,c & $ 9.3770 \pm 0.0001$ & $ 782.572 \pm 0.001$ & $-0.020^{+0.008}_{-0.009}$ & $-0.015^{+0.009}_{-0.010}$ &  &  & $ 3.6 \pm 0.2$   \\
\hline
\multicolumn{8}{|c|}{3-body Case~B (87 samples) $\min\chi^2$ = $696.778$ } \\
\hline
Kep-50\,b & $ 7.8121 \pm 0.0001 $ & $ 785.304^{+0.001}_{-0.002}$ & $ 0.027^{+0.004}_{-0.005}$ & $ 0.017 \pm 0.005 $ & {$-0.049^{+0.001}_{-0.002}$} & {$-0.033 \pm 0.002$} & $ 1.9 \pm 0.3$ \\
Kep-50\,c & $ 9.3764 \pm 0.0002 $ & $ 782.572 \pm 0.001$ & $-0.022^{+0.005}_{-0.004}$ & $-0.016^{+0.006}_{-0.005}$ &  &  & $ 3.5 \pm 0.2$   \\
\hline
\end{tabular}
\caption{Confidence intervals (16th/50th/84th-percentile) for the orbital parameters of Kepler-50 b and Kepler-50 c within three-planet systems for four sample selections, and the lowest $\chi^2$ value in each sub-sample. (1)~All 700,000 posterior samples. (2)~The 285,138 samples with $bc$ 6:5 libration and Pr(stable) $>$ 0.5 (green in Figures~\ref{Fig:resKDEs1} and \ref{Fig:resKDEs2}, and the nominal result of this study. (3) (Case~A)~The 208  samples that additionally satisfy 3-body+$cx$ 1st-order libration and Pr(stable) $>$ 0.5 (blue in Figure~\ref{Fig:resKDEs1}) (4) (Case~B)~The 87 samples with $cx$ 2nd-order libration (purple in Figure~\ref{Fig:resKDEs2}). The mid-transit time $T_{0}$ is in BJD-2,454,900.}\label{Tab:samples}
  \end{center}
\end{table*}

\subsection{Robust masses for Kepler-50 b and c.}
\citet{Lammers2026} found that the calculated masses of transiting planets could differ substantially between solutions where a non-transiting perturber cannot be uniquely characterized. However, in the case of Kepler-50 b and c, we obtain robust mass posteriors. 

For Kepler-50, the non-transiting planet improves the fit of the TTVs substantially by providing a long period TTV signal, but it is too low in mass to cause any detectable high-frequency, low-amplitude TTVs (known as `synodic chopping'). For example, using the solution in Table~\ref{Tab:mcmc} where $P_{x} = 12.4802$ days, we estimated the expected non-resonant TTV S/N given the median transit timing uncertainties and the empirical fit for the minimum non-resonant TTV given orbital periods and masses from \citet{Jontof-Hutter2021}. We found that non-resonant TTVs in Kepler-50 c are expected at the 6.5$\sigma$ level due to perturbations from Kepler-50 b, but only at the 0.17$\sigma$ level due to Kepler-50x. This prevents Kepler-50 x from contributing to the detected high frequency signal between Kepler-50 b and c and makes our models less susceptible to overfitting. 

By contrast, the discrepant mass estimates for the transiting planets in systems like Kepler-82 are likely the result of overfitting to some extent, whereby the model fits cannot decouple the synodic chopping signals from multiple perturbers. We test this explanation by estimating the chopping S/N between transiting planets of the sample in \cite{Lammers2026} for their degenerate solutions with disparate masses for the transiting planets of Kepler-82. We find the expected chopping S/N for Kepler-82 c at 130$\sigma$ due to the transiting planet Kepler-82 b and 67$\sigma$ due to the non-transiting perturber. Hence, high frequency TTVs are more susceptible to over-fitting from the putative planet in Kepler-82, leaving the mass estimates of the transiting planet degenerate.   

\begin{figure*}
\includegraphics[width = 3.5 in]{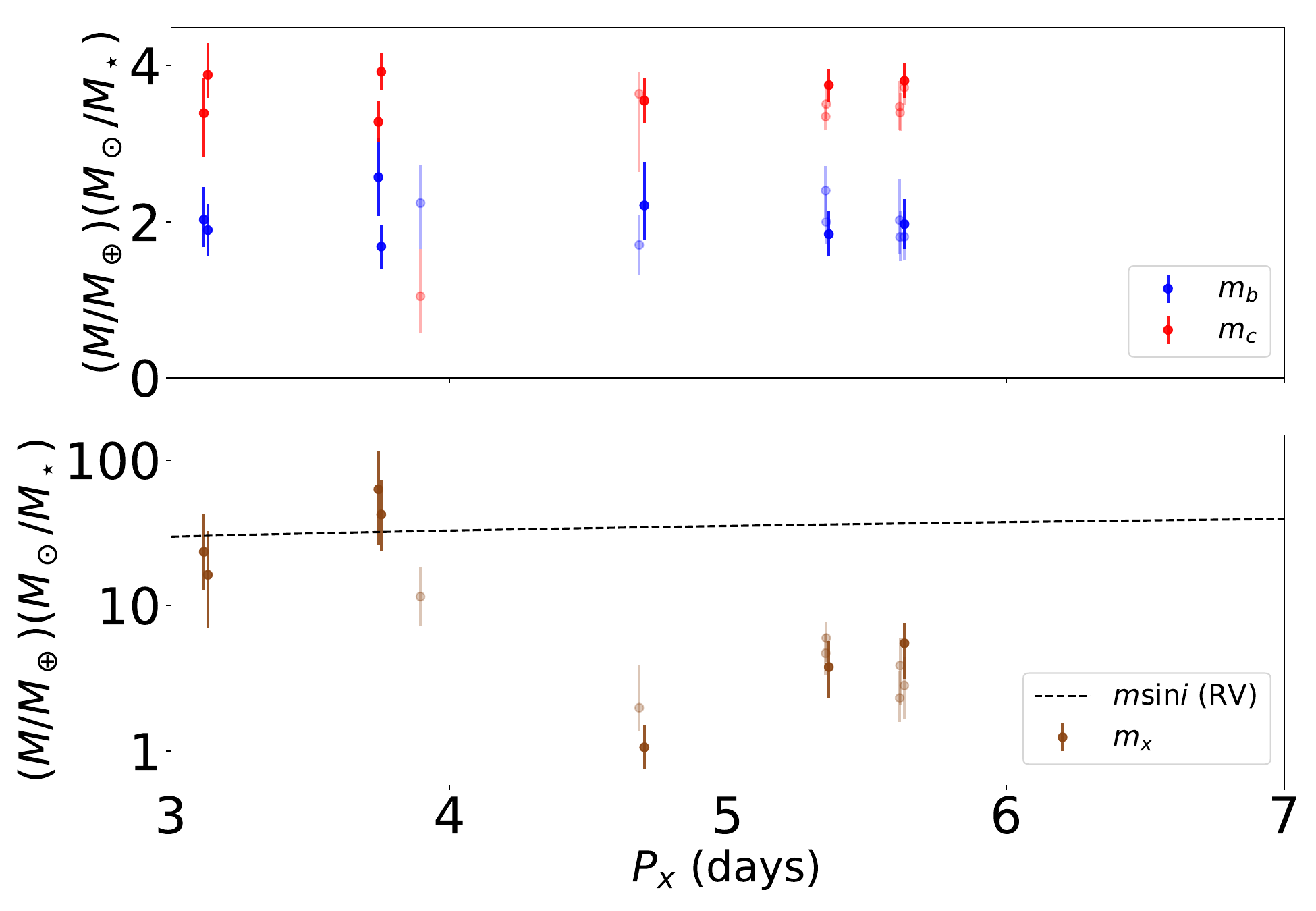}
\includegraphics[width = 3.5 in]{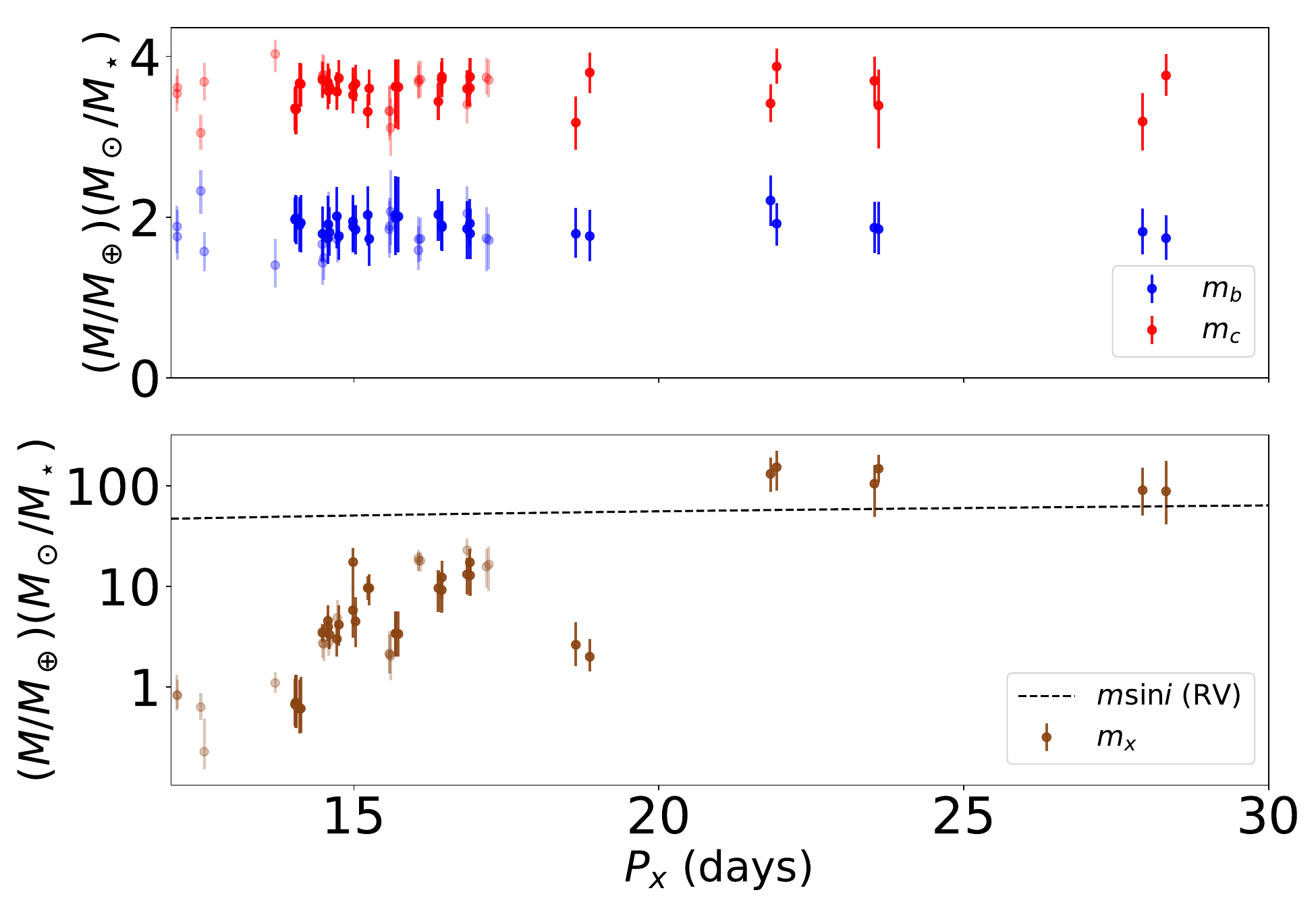}
\caption{The dynamical mass posteriors for b and c (top panels) and x (lower panels) where Pr(stable) $>$ 0.5. We display each mass posterior for sub-samples only at the region of interest identified via LM-minimization (see Section~\ref{sec:LMfits}), while it is the union of these sub-samples that we use for our nominal result. These regions of interest cover a wide range of 3-planet solutions, and indicate a robust mass measurement of the transiting planets, while posteriors for Kepler-50x vary significantly between regions of interest. Approximate upper limits from RV constraints adjusted for the stellar mass ($\delta \rm{RV} \lesssim 12$ m/s, \citealt{Weiss2024}) are shown as dashed curves. Highlighted are posteriors from the regions of interest where at least 80\% of samples are likely stable (Pr(stable) $>$ 0.5) and where at least 80\% of samples are in 2-body libration. The remaining regions are shown with fainter symbols.}
\label{Fig:MassPx}
\end{figure*}
As shown in Figure~\ref{Fig:MassPx}, the dynamical mass estimates of the transiting planets are remarkably consistent across a wide range of 3-planet solutions. We estimated the RV upper limits given the spectroscopic precision and non-detection reported by \cite{Weiss2024} for Kepler-50 ($\delta \rm{RV} \lesssim 12$ m/s) and the stellar mass measured by \citet{Chaplin2013} (see Section \ref{sec:FullyCharacterize}). The RV constraints do not rule out any region of interest, although solutions beyond 20 days appear less likely.

Given the masses that we have found via posterior sampling, we revisit the choices for the two-planet models that we used to initialize our three-planet grid search for regions of interest. While fixing the masses of the transiting planets at 3.5 M$_{\oplus}$ results in a $\chi^2 = 1157.6$, adopting the nominal measured masses from the three-planet fits for a two-planet model gives a $\chi^2 = 1062.5$, an improvement on our initial estimate for the masses but well short of the improvement provided by a three-planet model.

\section{Characterizing Kepler-50 \lowercase{b} and \lowercase{c}} \label{sec:FullyCharacterize}
The stellar properties of Kepler-50 have benefited from high-resolution spectra as well as strongly detected asteroseismic modes. \citet{Chaplin2013} characterized the star Kepler-50, using its asteroseismic power spectrum obtained from the Kepler light curve, which puts a tight constraint on the bulk density of the star. We list these stellar parameters in Table~\ref{Tab:star}. 
\begin{table}
  \begin{center}
  \begin{tabular}{|l|l|}
 \hline
Param & Value \\
\hline
$T_{\rm{eff}}$  (K) &  $6225 \pm 66$  \\
$[\rm{Fe/H}]$ (dex) & $0.03 \pm 0.06$ \\
$M$ (M$_{\odot}$) & $1.24 \pm 0.05$  \\
$R$ (R$_{\odot}$) & $1.58 \pm 0.02$  \\
$\langle \rho_{\star} \rangle$ (g cm$^{-3}$) & $0.441 \pm 0.004$  \\
log$g$ (dex) & $4.132 \pm 0.005$   \\
Age (Gyr) & $3.8 \pm 0.8$  \\
$P_{\rm rot}$ (days) & $7.7^{+0.5}_{-0.4}$   \\
\hline
\end{tabular}
\caption{Stellar parameters for Kepler-50 \citep{Chaplin2013}.}\label{Tab:star}
\end{center}
\end{table}

We used this precise measure of the stellar density as a prior for the transit model of the planets given 
\begin{equation} 
\rho_\star + \left(\frac{R_{p}}{R_{\star}}\right)^3 \rho_{p} = \frac{3\pi}{G P^2} \left( \frac{a}{R_\star} \right)^3,
\label{rhostar}
\end{equation}
(\citealt{Seager2003,Winn2010}). Here $\rho_{\star}$ is the bulk density of the star, $G$ is the gravitational constant, and $P$ is the orbital period of the transiting planet.
For small planets, the left-hand side reduces to $\rho_\star$. 

Furthermore, the eccentricity posteriors of our dynamical fits, alongside $\rho_\star$, permit greater precision in measuring the fractional transit depth $\delta$, the transit duration $T_{dur}$, and either the ingress or egress duration $\tau$ (following \citealt{Lissauer2013,Jontof-Hutter2014}):
\begin{equation}
\frac{a}{R_\star} = \frac{\delta^{1/4}}{\pi}\frac{P}{\sqrt{T\tau}}\left( \frac{\sqrt{1-e^2}}{1+e\sin\omega}\right)
\label{arstar}
\end{equation}
\citep{Winn2010}. 

We fit for stellar limb darkening coefficients using MARCS stellar atmosphere models \citep{Gustafsson2008}; $q_{1} = 0.376 \pm 0.006 $ and $q_{2} = 0.270 \pm 0.012$. The revised light curve parameters for the transiting planets are in Table~\ref{Tab:revisedLC}, and they differ slightly from the results of \citet{Lissauer2024}. Our calculated value of the radius of Kepler-50 b is almost identical but with tighter uncertainties, while the radius of Kepler-50 c is revised upward, to the sub-Neptune side of the `radius valley'. 

\begin{table}
  \begin{center}
  \begin{tabular}{|c|l|l|}
 \hline
Param & Kepler-50 b & Kepler-50 c \\
\hline
$b$ & 0.646$^{+0.024}_{-0.018 }$ & 0.9395$^{+0.0042}_{-0.0045}$ \\
$R_{p}$/R$_{\star}$ & 0.01016 $^{+ 0.00017}_{-0.00018}$ & 0.01200$^{+0.00037}_{-0.00025}$ \\
$a/R_\star$ & 11.252$^{+  0.032}_{-0.036}$ & 12.707 $^{+0.036}_{-0.040}$ \\
$\delta$ (ppm) & 110.8 $^{+ 3.3}_{-3.6}$ & 111.7$^{+6.9}_{-3.4}$ \\  
$T_{dur}$ (h) & 4.070$^{+ 0.083}_{-0.095}$ & 2.158$^{+   0.064}_{-0.063}$ \\ 
$R_{p}$ (R$_\oplus$) & 1.750 $^{+ 0.038}_{-0.037}$ & 2.079 $^{+ 0.060}_{-0.060}$ \\
Inclination (deg)  & 86.63$^{+0.15}_{-0.16}$ & 85.80$\pm 0.24$ \\ 
\hline
\end{tabular}
\caption{Revised light curve parameters for Kepler-50 b and Kepler-50 c using dynamical constraints on eccentricity vectors as priors.}\label{Tab:revisedLC}
\end{center}
\end{table}

\begin{figure*}
\includegraphics[width = 6.5 in]{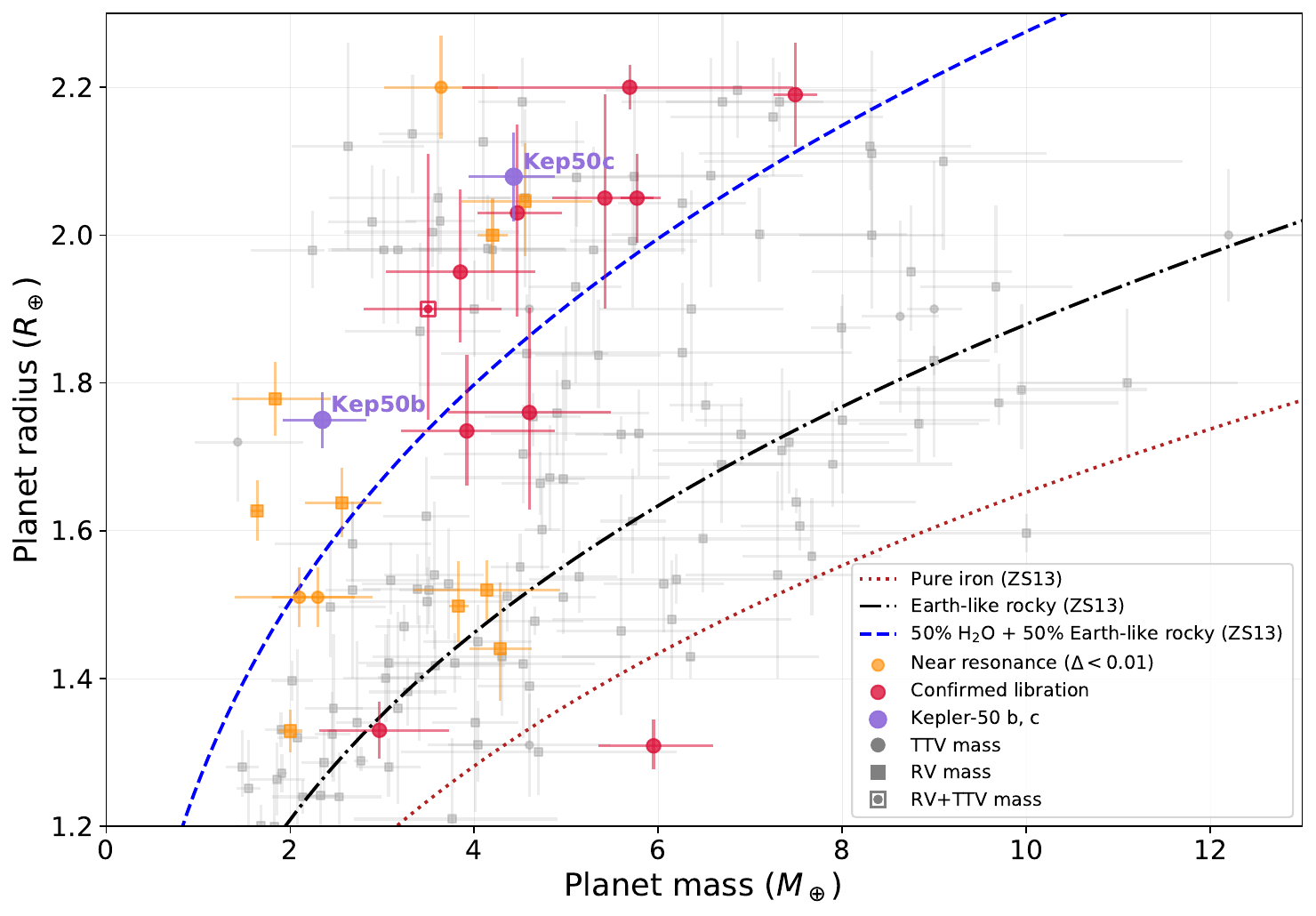 }
\caption{Mass-Radius diagram across the `radius valley', with Kepler-50 b and c (in purple) joining a sample of planets known to be in resonance (red) or near resonance (orange). We mark theoretical curves of iron (dotted red curve), an Earth-like mixture of iron and rock (black, dot-dash curve), and a `water-world' solid mix of water ice, rock and iron (blue dashed curve) (ZS13, \citealt{Zeng2013}). We distinguish TTV mass measurements (circles) from RV (squares). We limit the diagram to mass and radius detections where $M_{p}/\sigma_{M_{p}} > 3$ and $R_{p}/\sigma_{R_{p}} > 3$, and list our sources in Tables~\ref{tab:MRdata} and \ref{tab:MRdata2}.}
\label{Fig:MRdiagram}
\end{figure*}

The confidence intervals of the masses and bulk densities of the transiting planets are summarized in Table~\ref{tab:results}.
\begin{table}[h!]
\centering
\caption{Kepler-50 planet physical properties derived from TTV samples where $\mathrm{Pr(stable)}>0.5$ and that are likely in a $bc$ 2-body resonance.}
\label{tab:results}
\begin{tabular}{lccc}
\hline
Planet & $R_p$ ($R_\oplus$) & $M_p$ ($M_\oplus$) & $\rho_p$ (g\,cm$^{-3}$)  \\
\hline
Kepler-50\,b & $1.750^{+0.038}_{-0.038}$ & $2.35^{+0.48}_{-0.43}$ & $2.42^{+0.53}_{-0.46}$ \\
Kepler-50\,c & $2.079^{+0.060}_{-0.060}$ & $4.43^{+0.45}_{-0.49}$ & $2.71^{+0.39}_{-0.37}$ \\
\hline
\end{tabular}
\end{table}
Given the robustness of our mass measurements for Kepler-50 b and c, we place them on the mass-radius diagram in Figure~\ref{Fig:MRdiagram}. 

\section{Discussion and Conclusions}
A two-planet model for the transit times of Kepler-50 b and c leaves a significant residual signal that can be fit with a missing planet. We searched for regions with solutions for a third planet using LM fits over a grid of fixed periods with the masses of the transiting planets fixed at 3.5 M$_{\oplus}$. We identified regions of interest where stable solutions were plausible and explored each region of interest with posterior sampling.

Although the period and mass of Kepler-50x is degenerate, we strongly detect plausible and well-constrained masses for Kepler-50 b and c with 3-planet dynamical model fits to the observed transit times. In addition, we strongly detect non-zero relative eccentricities in the transiting planets. 

\cite{Leleu2024} discovered that sub-Neptunes in resonance are puffier than non-resonant planets in the radius range from 2--4 R$_{\oplus}$. Kepler-50 b is on the smaller end of the radius range considered by \citet{Leleu2024} and sits in the `radius valley', occupying a space in planetary radius noted for its relative paucity, possibly due to a transition from volatile-poor super-Earths to sub-Neptunes with deep atmospheres \citep{Fulton2017}, as well as the sensitivity of a low-mass planet's radius to a small fraction of its mass in the form of a gaseous atmosphere \citep{Lopez2014}. Kepler-50 c has a radius that places it among sub-Neptunes. Both planets are less dense than a `water world' mix of ice and rock following the models of \citet{Zeng2013} and likely require a deep envelope. 

As noted by \citet{Leleu2024}, the observed excess of volatile-rich sub-Neptunes in resonance must be disentangled from the known detection biases in the mass-radius diagram. TTV detections are biased toward lower densities since, for a given mass, TTVs are biased toward larger radii (more precisely measured transit times) and hence lower densities (\citealt{Steffen2016,Mills2017a}). Furthermore, orbital period ratios near resonances make a TTV detection much more likely. By contrast, RV detections of transiting planets are insensitive to resonance and scale with mass, favoring higher densities \citep{Steffen2016}.   

We note in Figure~\ref{Fig:MRdiagram} that below the radius valley, two planets around 1.3 $R_{\oplus}$ are in resonance (Kepler-80 d and e). However, across the radius valley from 1.5--2 $R_{\oplus}$, none of the 18 planets that are nominally denser than rocky are in resonance or near enough to resonance such that $\Delta < 0.01$. A Fisher Exact Test of these groups including RV and TTV mass measurements, and keeping the resonant and near-resonant planets as one group, gives a p-value = 0.034. This is the likelihood that density relative to a rocky planet has no effect on whether a planet is in resonance. If we restrict ourselves to the sample known to be in libration, the p-value = 0.264. However, if we use the RV sample alone across the radius valley, there is only one planet characterized by RV in the radius range that is known to be in libration (TOI-1136 b, which is detected with both RV and TTV; \citealt{Beard2024}). This sub-sample has no statistical power to detect an association between libration and composition (p-value = 0.757). Finally, if we combine known resonant and near-resonant planets in the RV sample, we find p = 0.129. Hence, there are not enough RV-only mass detections in the radius valley regime to show a strong preference for lower density if planets are in resonance. We conclude here that the excess of low-density planets in resonance across the radius valley --- a sample which now includes Kepler-50 b --- is consistent with the known detection bias of TTVs toward lower-density planets and toward systems with period ratios near mean motion resonance, and therefore a preference for volatile retention in resonant systems in this radius range remains unconfirmed.

Although the fraction of our TTV posteriors where the three planets of Kepler-50 occupy a resonance chain is small, such an outcome is not ruled out by the data. Water-rich sub-Neptunes most likely formed beyond the snowline before migrating to their observed locations near the star \citep{Burn2024}. \cite{Dai2024} found a decline in the frequency of resonance chains with age, though it is unclear what the implications are for the presence of 2-body librations only. We note that Kepler-50 is a mature system, with age $\approx 3.8$ Gyr according to \citet{Chaplin2013}.  

In summary, Kepler-50 presents a rare case in which a non-transiting perturber with poorly-defined (degenerate) mass and orbit solutions improves rather than degrades the inferred posteriors on the mass of the transiting planets. The robustness of the mass posteriors for Kepler-50 b and c across 70 distinct solutions for the third planet's orbital period demonstrates that high-frequency TTVs (synodic chopping) are detectable between the planets but undetectable from the perturber which is either low in mass or at a relatively large period ratio. The resulting masses are consistent with bulk compositions that include the retention of a deep atmosphere.

The strongly detected non-zero relative eccentricity implies anti-aligned orbital pericenters for Kepler-50 b and c. We note that the 6:5 resonant state of the pair cannot be confirmed by TTVs alone since $\approx 40\%$ of posterior samples (68\% of the stable samples) librate. Although a small fraction of samples and a higher $\chi^2$ makes a 3-body resonance configuration unlikely, a resonance chain remains at least plausible. 

\begin{acknowledgments}
The authors thank C. Lammers for discussions that improved this paper.
\end{acknowledgments}

\begin{contribution}
DJ was responsible for leading the analysis of the transit timing variations with dynamical models, assessing the stability of solutions, identifying resonances and writing the manuscript.
JR analyzed the Kepler data to measure transit times, and used posteriors from the dynamical models to revisit the light curve parameters. 
JL and DF contributed to the analysis methodology of 3-planet models and interpretation of the results. 
\end{contribution}

\clearpage
\appendix
\begin{table}[h]
\tiny
\centering
\caption{Kepler-50 transit times (BJD-2454900) and uncertainties.}
\label{tab:TT}
\begin{tabular}{c@{\hskip 0.5\tabcolsep}c@{\hskip 2\tabcolsep}c@{\hskip 0.5\tabcolsep}c@{\hskip 2\tabcolsep}c@{\hskip 0.5\tabcolsep}c@{\hskip 2\tabcolsep}c@{\hskip 0.5\tabcolsep}c@{\hskip 2\tabcolsep}c@{\hskip 0.5\tabcolsep}c@{\hskip 2\tabcolsep}c@{\hskip 0.5\tabcolsep}c@{\hskip 2\tabcolsep}c@{\hskip 0.5\tabcolsep}c}
\hline
$t$  & $\sigma_t$ & $t$  & $\sigma_t$ & $t$  & $\sigma_t$ & $t$  & $\sigma_t$ & $t$  & $\sigma_t$ & $t$  & $\sigma_t$ & $t$  & $\sigma_t$ \\
\hline
58.76690 & 0.00776 & 441.59133 & 0.01253 & 816.53323 & 0.01139 & 1215.04097 & 0.00608 & 201.23525 & 0.00544 & 632.53245 & 0.00621 & 1091.99390 & 0.00974 \\
66.51472 & 0.01081 & 449.39665 & 0.00759 & 824.39002 & 0.00935 & 1230.67426 & 0.01429 & 210.61454 & 0.01876 & 641.89805 & 0.00604 & 1101.36845 & 0.00946 \\
74.36238 & 0.01199 & 457.22666 & 0.02417 & 832.17295 & 0.02910 & 1246.30355 & 0.01767 & 219.97482 & 0.01274 & 651.30409 & 0.02503 & 1110.75371 & 0.01857 \\
82.19125 & 0.01478 & 465.01493 & 0.00880 & 839.98598 & 0.01462 & 1254.16008 & 0.00669 & 229.35408 & 0.00659 & 670.05785 & 0.01013 & 1120.14409 & 0.00394 \\
89.97461 & 0.01745 & 472.81779 & 0.00727 & 847.77537 & 0.01711 & 1261.92417 & 0.03066 & 238.73636 & 0.01571 & 679.44021 & 0.03611 & 1129.51135 & 0.01648 \\
97.81037 & 0.01183 & 480.65141 & 0.01075 & 855.63960 & 0.00997 & 1277.55499 & 0.02544 & 248.12463 & 0.01448 & 688.80255 & 0.01836 & 1138.87945 & 0.01352 \\
105.60084 & 0.01834 & 488.46735 & 0.01141 & 863.45682 & 0.03138 & 1285.37802 & 0.00645 & 257.48086 & 0.01241 & 698.17990 & 0.01308 & 1157.63987 & 0.03756 \\
113.45702 & 0.00708 & 496.25813 & 0.01897 & 871.22936 & 0.00522 & 1293.16574 & 0.01341 & 266.86382 & 0.01498 & 707.55641 & 0.02626 & 1167.00635 & 0.01937 \\
121.24017 & 0.00787 & 504.08897 & 0.01288 & 879.03222 & 0.01202 & 1300.96976 & 0.00723 & 276.25695 & 0.00704 & 716.94122 & 0.02199 & 1176.31831 & 0.03048 \\
129.07039 & 0.01302 & 511.88540 & 0.00474 & 886.88761 & 0.01600 & 1308.83088 & 0.00459 & 285.62735 & 0.02095 & 726.29884 & 0.00682 & 1185.75765 & 0.01653 \\
136.86984 & 0.00613 & 519.71353 & 0.00659 & 894.70833 & 0.01620 & 1316.61885 & 0.00474 & 294.99995 & 0.01290 & 745.07239 & 0.00537 & 1195.13863 & 0.00766 \\
144.68020 & 0.01239 & 527.50507 & 0.00916 & 910.29870 & 0.00788 & 1324.43174 & 0.01180 & 304.35994 & 0.01861 & 754.43935 & 0.01012 & 1204.51966 & 0.00873 \\
152.50662 & 0.00721 & 535.32632 & 0.00688 & 918.11575 & 0.01818 & 1332.27255 & 0.01229 & 313.75327 & 0.03177 & 763.81794 & 0.01024 & 1213.88533 & 0.00748 \\
160.30336 & 0.02501 & 543.14507 & 0.00993 & 925.94587 & 0.00904 & 1340.07315 & 0.01986 & 323.11646 & 0.00992 & 773.21076 & 0.01510 & 1232.64662 & 0.00709 \\
168.11518 & 0.02242 & 550.94806 & 0.00967 & 941.56363 & 0.01292 & 1355.75198 & 0.00797 & 341.87782 & 0.03027 & 782.57538 & 0.01278 & 1242.04113 & 0.01616 \\
175.94015 & 0.01183 & 558.77343 & 0.02101 & 949.35270 & 0.01472 & 1363.48877 & 0.02099 & 351.24657 & 0.00682 & 791.96477 & 0.00507 & 1251.40027 & 0.01446 \\
183.75562 & 0.00743 & 566.56843 & 0.01461 & 957.21118 & 0.02814 & 1371.30222 & 0.00426 & 360.62709 & 0.02603 & 801.33965 & 0.02070 & 1260.78151 & 0.01153 \\
199.38580 & 0.00677 & 574.39890 & 0.01257 & 965.04586 & 0.01519 & 1379.12755 & 0.01406 & 369.99106 & 0.01481 & 810.70371 & 0.00548 & 1279.54278 & 0.01725 \\
207.20950 & 0.01137 & 582.17448 & 0.01437 & 972.81825 & 0.01239 & 1386.93692 & 0.01308 & 379.39063 & 0.00786 & 829.46288 & 0.00947 & 1288.91863 & 0.02518 \\
214.98610 & 0.01607 & 589.98256 & 0.00990 & 980.61960 & 0.02428 & 1394.73849 & 0.01551 & 388.73721 & 0.00396 & 838.83726 & 0.00594 & 1298.28502 & 0.01050 \\
222.80727 & 0.00668 & 597.81022 & 0.02877 & 988.45252 & 0.02836 & 1402.54068 & 0.02123 & 398.13497 & 0.01450 & 848.21353 & 0.01416 & 1307.67067 & 0.00916 \\
230.67881 & 0.01057 & 605.62635 & 0.04010 & 1011.88631 & 0.02032 & 1425.99893 & 0.00864 & 407.50899 & 0.01648 & 857.59068 & 0.01942 & 1317.04781 & 0.01197 \\
238.43142 & 0.02627 & 613.43949 & 0.00972 & 1019.72356 & 0.00905 & 1433.82167 & 0.01247 & 416.88549 & 0.01409 & 866.97316 & 0.01959 & 1326.42440 & 0.01078 \\
246.26598 & 0.05013 & 621.25017 & 0.01000 & 1027.50961 & 0.02011 & 1441.62523 & 0.01235 & 426.25993 & 0.04883 & 876.33680 & 0.01463 & 1335.80683 & 0.00745 \\
254.05464 & 0.00842 & 629.06928 & 0.01101 & 1035.36386 & 0.02385 & 1449.43857 & 0.01484 & 435.65214 & 0.01323 & 885.71912 & 0.00709 & 1345.17850 & 0.01106 \\
261.88455 & 0.01992 & 636.89840 & 0.01320 & 1043.14488 & 0.01316 & 1457.25597 & 0.02127 & 445.01838 & 0.00541 & 895.09743 & 0.00650 & 1354.55699 & 0.01211 \\
269.72696 & 0.02125 & 644.69654 & 0.01227 & 1058.76575 & 0.01004 & 1465.05415 & 0.01400 & 454.39229 & 0.00442 & 904.46135 & 0.00879 & 1363.93471 & 0.00919 \\
277.48265 & 0.01251 & 652.49982 & 0.01785 & 1066.50918 & 0.00773 & 1472.85719 & 0.00552 & 463.76155 & 0.00660 & 913.84884 & 0.00790 & 1373.31183 & 0.00710 \\
293.15872 & 0.01087 & 675.93771 & 0.00845 & 1074.38708 & 0.01202 & 1480.69072 & 0.01412 & 473.13361 & 0.00985 & 923.21948 & 0.00603 & 1382.68806 & 0.00763 \\
300.93988 & 0.06441 & 683.72721 & 0.00964 & 1082.22397 & 0.01168 & 1488.50668 & 0.01160 & 482.50150 & 0.02671 & 932.60297 & 0.00870 & 1392.10050 & 0.01964 \\
308.75462 & 0.00879 & 691.55449 & 0.02226 & 1090.03341 & 0.01809 & 1496.33842 & 0.00651 & 491.89310 & 0.00502 & 941.98785 & 0.01548 & 1401.43642 & 0.01438 \\
324.33025 & 0.00988 & 699.40304 & 0.00919 & 1105.62808 & 0.01222 & 1504.11866 & 0.00813 & 501.27676 & 0.01295 & 951.35672 & 0.01147 & 1429.58324 & 0.01166 \\
340.00132 & 0.00957 & 707.20652 & 0.00773 & 1113.47548 & 0.01634 & 1511.97014 & 0.03155 & 510.65239 & 0.01070 & 960.73604 & 0.01925 & 1438.96166 & 0.00689 \\
347.82178 & 0.02750 & 715.01083 & 0.00671 & 1121.26906 & 0.00865 & 60.56350 & 0.01664 & 520.02014 & 0.01137 & 970.11175 & 0.02012 & 1448.33967 & 0.01830 \\
355.66382 & 0.00950 & 722.79928 & 0.01905 & 1129.08785 & 0.04029 & 69.97606 & 0.00417 & 529.39853 & 0.00734 & 979.49036 & 0.00838 & 1457.71684 & 0.01048 \\
363.43686 & 0.03265 & 730.61955 & 0.01987 & 1136.90970 & 0.02982 & 79.33032 & 0.00428 & 538.76255 & 0.00486 & 988.85259 & 0.02964 & 1467.07761 & 0.01248 \\
371.24681 & 0.03131 & 746.23135 & 0.00799 & 1144.72252 & 0.02327 & 88.72249 & 0.01281 & 548.15143 & 0.01303 & 998.23290 & 0.00958 & 1476.46958 & 0.02039 \\
379.05708 & 0.01729 & 754.04514 & 0.00760 & 1152.55915 & 0.02521 & 107.48399 & 0.01240 & 557.53512 & 0.01246 & 1007.61329 & 0.06164 & 1485.84347 & 0.01451 \\
386.91220 & 0.01043 & 761.86987 & 0.00689 & 1160.35255 & 0.01885 & 116.85968 & 0.01261 & 566.85262 & 0.01755 & 1016.99154 & 0.02171 & 1495.21937 & 0.00764 \\
394.69415 & 0.02382 & 769.68983 & 0.00775 & 1168.16054 & 0.01114 & 126.23218 & 0.00776 & 576.28880 & 0.01043 & 1026.36509 & 0.00449 & 1504.59584 & 0.00669 \\
402.51189 & 0.02282 & 777.48553 & 0.01642 & 1175.97180 & 0.07434 & 135.60485 & 0.00635 & 585.64875 & 0.01755 & 1035.74950 & 0.02692 & 1513.97405 & 0.00729 \\
410.36112 & 0.02832 & 785.28968 & 0.01262 & 1183.78224 & 0.01958 & 144.98248 & 0.00849 & 595.03382 & 0.00753 & 1045.11059 & 0.01995 & 1523.33003 & 0.01169 \\
418.14808 & 0.02877 & 793.11048 & 0.00832 & 1191.59550 & 0.03709 & 154.36572 & 0.01102 & 604.40911 & 0.01341 & 1063.87560 & 0.00994 &  &  \\
425.92425 & 0.02607 & 800.92548 & 0.03371 & 1199.41677 & 0.01131 & 173.09315 & 0.00845 & 613.77859 & 0.00619 & 1073.25248 & 0.01061 &  &  \\
433.76702 & 0.03229 & 808.73682 & 0.03436 & 1207.23430 & 0.01436 & 182.49374 & 0.00771 & 623.16305 & 0.04623 & 1082.61898 & 0.01535 &  &  \\
\end{tabular}
\end{table}

\begin{table}
\tiny
\begin{center}
\caption{Planetary mass and radius references (Part 1 of 2). \label{tab:MRdata}}
\begin{tabular}{llll|llll}
Planet & $M_p$ ($M_\oplus$) & $R_p$ ($R_\oplus$) & Ref. &
Planet & $M_p$ ($M_\oplus$) & $R_p$ ($R_\oplus$) & Ref. \\
\hline
 55 Cnc e & $7.99^{+0.32}_{-0.33}$ & $1.875 \pm 0.029$ & \cite{Bourrier2018} &  Kepler-80 e & $2.97^{+0.76}_{-0.65}$ & $1.330^{+0.039}_{-0.038}$ & \cite{MacDonald2021} \\
 CoRoT-7 b & $6.06 \pm 0.65$ & $1.528 \pm 0.065$ & \cite{John2022} &  Kepler-85 b & $1.84^{+0.60}_{-0.47}$ & $1.778 \pm 0.050$ & \cite{Leleu2023} \\
  G 9-40 b & $4.00 \pm 0.63$ & $1.900 \pm 0.065$ & \cite{Luque2022c} &  Kepler-93 b & $4.66 \pm 0.53$ & $1.478 \pm 0.019$ & \cite{Dressing2015} \\
  GJ 3090 b & $4.52 \pm 0.47$ & $2.18 \pm 0.06$ & \cite{Almenara2022} &  Kepler-99 b & $6.15 \pm 1.30$ & $1.48 \pm 0.08$ & \cite{Marcy2014} \\
  GJ 3473 b & $1.86 \pm 0.30$ & $1.264^{+0.050}_{-0.049}$ & \cite{Kemmer2020} &  L 168-9 b & $4.60 \pm 0.56$ & $1.39 \pm 0.09$ & \cite{Alam2025} \\
  GJ 357 b & $1.83 \pm 0.30$ & $1.20 \pm 0.06$ & \cite{Luque2019} &  L 98-59 c & $2.00 \pm 0.13$ & $1.329 \pm 0.029$ & \cite{Cadieux2025} \\
  GJ 486 b & $2.770^{+0.076}_{-0.073}$ & $1.289^{+0.019}_{-0.014}$ & \cite{Trifonov2021} &  L 98-59 d & $1.64 \pm 0.07$ & $1.627 \pm 0.041$ & \cite{Cadieux2025} \\
   GJ 806 b & $1.90 \pm 0.17$ & $1.331 \pm 0.023$ & \cite{Palle2023} &  LHS 1140 b & $5.60 \pm 0.19$ & $1.730 \pm 0.025$ & \cite{Cadieux2024} \\
  GJ 9827 b & $4.28^{+0.35}_{-0.33}$ & $1.44^{+0.09}_{-0.07}$ & \cite{Passegger2024} &  LHS 1140 c & $1.91 \pm 0.06$ & $1.272 \pm 0.026$ & \cite{Cadieux2024} \\
  GJ 9827 d & $3.02^{+0.58}_{-0.57}$ & $1.98^{+0.11}_{-0.10}$ & \cite{Passegger2024} &  LHS 1478 b & $2.33 \pm 0.20$ & $1.242^{+0.051}_{-0.049}$ & \cite{Soto2021} \\
  \hline
\end{tabular}
\end{center}
\end{table}

\begin{table}
\tiny
\begin{center}
\caption{Planetary mass and radius references (Part 2 of 2). Radius updates from \citet{Lissauer2024} are labelled L24. \label{tab:MRdata2}}
\begin{tabular}{llll|llll}
Planet & $M_p$ ($M_\oplus$) & $R_p$ ($R_\oplus$) & Ref. &
Planet & $M_p$ ($M_\oplus$) & $R_p$ ($R_\oplus$) & Ref. \\
\hline
   HD 110067 b & $5.69^{+1.78}_{-1.82}$ & $2.20 \pm 0.03$ & \cite{Luque2023} &  LHS 1903 b & $3.28 \pm 0.42$ & $1.382 \pm 0.046$ & \cite{Wilson2026} \\
  HD 110113 b & $4.55 \pm 0.62$ & $2.05 \pm 0.12$ & \cite{Osborn2021b} &  LHS 1903 c & $4.55^{+0.73}_{-0.69}$ & $2.046^{+0.078}_{-0.074}$ & \cite{Wilson2026} \\
  HD 136352 b & $4.72 \pm 0.42$ & $1.664 \pm 0.043$ & \cite{Delrez2021} &  LHS 1903 e & $5.79^{+1.60}_{-1.61}$ & $1.732^{+0.059}_{-0.058}$ & \cite{Wilson2026} \\
  HD 137496 b & $4.04 \pm 0.55$ & $1.31^{+0.06}_{-0.05}$ & \cite{AzevedoSilva2022} &  LHS 3844 b & $2.37 \pm 0.25$ & $1.286^{+0.043}_{-0.044}$ & \cite{Nagel2026} \\
  HD 15337 b & $6.52^{+0.41}_{-0.40}$ & $1.770^{+0.032}_{-0.030}$ & \cite{Dumusque2019} &  LTT 1445 A b & $2.73^{+0.25}_{-0.23}$ & $1.34^{+0.11}_{-0.06}$ & \cite{Lavie2023} \\
  HD 20329 b & $7.42 \pm 1.09$ & $1.72 \pm 0.07$ & \cite{Murgas2022} &  LTT 3780 b & $2.46 \pm 0.19$ & $1.325^{+0.057}_{-0.058}$ & \cite{Cloutier2020} \\
  HD 213885 b & $8.83^{+0.66}_{-0.65}$ & $1.745^{+0.051}_{-0.052}$ & \cite{Espinoza2020} &  pi Men c & $3.63 \pm 0.38$ & $2.019^{+0.046}_{-0.045}$ & \cite{Hatzes2022} \\
  HD 219134 b & $4.74 \pm 0.19$ & $1.602 \pm 0.055$ & \cite{Gillon2017b} &  Ross 176 b & $4.57^{+0.89}_{-0.93}$ & $1.84 \pm 0.08$ & \cite{Geraldia-Gonzalez2025} \\
  HD 219134 c & $4.36 \pm 0.22$ & $1.511 \pm 0.047$ & \cite{Gillon2017b} &  TOI-1011 b & $4.04 \pm 0.59$ & $1.45 \pm 0.05$ & \cite{Brinkman2025} \\
  HD 23472 b & $8.32^{+0.78}_{-0.79}$ & $2.00^{+0.11}_{-0.10}$ & \cite{Barros2022} &  TOI-1075 b & $9.95^{+1.36}_{-1.30}$ & $1.79^{+0.12}_{-0.08}$ & \cite{Essack2023} \\
  HD 23472 c & $3.41^{+0.88}_{-0.81}$ & $1.87^{+0.12}_{-0.11}$ & \cite{Barros2022} &  TOI-1136 b & $3.5^{+0.8}_{-0.7}$ & $1.90^{+0.21}_{-0.15}$ & \cite{Beard2024} \\
  HD 260655 b & $2.14 \pm 0.34$ & $1.240 \pm 0.023$ & \cite{Luque2022b} &  TOI-1203 b & $3.51^{+0.33}_{-0.32}$ & $1.520^{+0.045}_{-0.046}$ & \cite{Gandolfi2025} \\
  HD 260655 c & $3.09 \pm 0.48$ & $1.533^{+0.051}_{-0.046}$ & \cite{Luque2022b} &  TOI-1235 b & $6.69^{+0.67}_{-0.69}$ & $1.69 \pm 0.08$ & \cite{Bluhm2020} \\
  HD 3167 b & $4.97^{+0.24}_{-0.23}$ & $1.67^{+0.17}_{-0.10}$ & \cite{Gandolfi2017} &  TOI-1238 b & $3.76^{+1.15}_{-1.07}$ & $1.21^{+0.11}_{-0.10}$ & \cite{Gonzalez-Alvarez2022} \\
  HD 80653 b & $5.72^{+0.36}_{-0.35}$ & $1.613 \pm 0.071$ & \cite{Frustagli2020} &  TOI-1238 c & $8.32^{+1.90}_{-1.88}$ & $2.11 \pm 0.14$ & \cite{Gonzalez-Alvarez2022} \\
  HD 86226 c & $7.25^{+1.19}_{-1.12}$ & $2.16 \pm 0.08$ & \cite{Teske2020} &  TOI-1266 c & $2.64 \pm 0.52$ & $2.13 \pm 0.12$ & \cite{Tyler2025} \\
  HD 97658 b & $8.3 \pm 1.1$ & $2.12 \pm 0.06$ & \cite{VanGrootel2014} &  TOI-1347 b & $11.1 \pm 1.2$ & $1.8 \pm 0.1$ & \cite{Rubenzahl2024} \\
  HIP 29442 c & $4.50 \pm 0.32$ & $1.551 \pm 0.045$ & \cite{Damasso2023} &  TOI-1416 b & $3.48 \pm 0.47$ & $1.62 \pm 0.08$ & \cite{Deeg2023} \\
  HIP 29442 d & $5.14 \pm 0.41$ & $1.538 \pm 0.049$ & \cite{Damasso2023} &  TOI-1430 b & $4.2 \pm 0.8$ & $1.98 \pm 0.07$ & \cite{Nardiello2025} \\
  K2-106 b & $7.54^{+0.65}_{-0.64}$ & $1.606^{+0.034}_{-0.033}$ & \cite{Livingston2024} &  TOI-1444 b & $3.58^{+0.76}_{-0.75}$ & $1.418^{+0.047}_{-0.041}$ & \cite{Dai2021} \\
  K2-111 b & $5.35^{+0.68}_{-0.69}$ & $1.837^{+0.076}_{-0.071}$ & \cite{Mortier2020} &  TOI-1452 b & $4.82 \pm 1.30$ & $1.672 \pm 0.071$ & \cite{Cadieux2022} \\
  K2-131 b & $7.9 \pm 1.3$ & $1.690^{+0.085}_{-0.058}$ & \cite{Dai2017} &  TOI-1468 b & $3.04 \pm 0.46$ & $1.401^{+0.059}_{-0.060}$ & \cite{Chaturvedi2022} \\
  K2-141 b & $4.97^{+0.35}_{-0.34}$ & $1.51 \pm 0.05$ & \cite{Barragan2018} &  TOI-1468 c & $4.1 \pm 1.1$ & $2.126^{+0.092}_{-0.093}$ & \cite{Chaturvedi2022} \\
  K2-146 b & $5.77 \pm 0.18$ & $2.05 \pm 0.06$ & \cite{Hamann2019} &  TOI-1470 b & $7.32^{+1.21}_{-1.24}$ & $2.18 \pm 0.04$ & \cite{Gonzalez-Alvarez2023} \\
  K2-146 c & $7.49 \pm 0.24$ & $2.19 \pm 0.07$ & \cite{Hamann2019} &  TOI-1634 b & $4.9 \pm 1.0$ & $1.759^{+0.058}_{-0.057}$ & \cite{Cloutier2021} \\
  K2-199 b & $6.9 \pm 1.8$ & $1.73^{+0.05}_{-0.04}$ & \cite{AkanaMurphy2021} &  TOI-1685 b & $3.07^{+0.34}_{-0.33}$ & $1.421 \pm 0.060$ & \cite{Burt2024} \\
  K2-216 b & $8.0 \pm 1.6$ & $1.75^{+0.17}_{-0.10}$ & \cite{Persson2018} &  TOI-1695 b & $6.36 \pm 1.00$ & $1.90^{+0.16}_{-0.14}$ & \cite{Kiefer2023} \\
  K2-265 b & $7.34^{+1.43}_{-1.40}$ & $1.71^{+0.11}_{-0.08}$ & \cite{Lam2018} &  TOI-178 c & $4.64^{+0.52}_{-0.53}$ & $1.754^{+0.032}_{-0.040}$ & \cite{Leleu2022} \\
  K2-291 b & $6.49 \pm 1.16$ & $1.589^{+0.095}_{-0.072}$ & \cite{Kosiarek2019b} &  TOI-1798.02 & $5.6^{+0.8}_{-0.7}$ & $1.46^{+0.18}_{-0.11}$ & \cite{Polanski2024} \\
  K2-3 b & $5.11^{+0.65}_{-0.64}$ & $2.078^{+0.076}_{-0.067}$ & \cite{Kosiarek2019a} &  TOI-1801 b & $5.74 \pm 1.46$ & $2.08^{+0.12}_{-0.11}$ & \cite{Mallorquin2023} \\
  K2-3 c & $2.68 \pm 0.85$ & $1.582^{+0.057}_{-0.051}$ & \cite{Diamond-Lowe2022} &  TOI-1807 b & $2.44^{+0.60}_{-0.57}$ & $1.497^{+0.081}_{-0.068}$ & \cite{Nardiello2022} \\
  K2-314 b & $8.75^{+1.09}_{-1.08}$ & $1.95^{+0.09}_{-0.08}$ & \cite{Hidalgo2020} &  TOI-198 b & $3.17^{+0.64}_{-0.65}$ & $1.36 \pm 0.13$ & \cite{ZapateroOsorio2026} \\
  K2-36 b & $4.3 \pm 1.4$ & $1.43 \pm 0.08$ & \cite{Bonomo2023} &  TOI-2076 e & $4.7 \pm 1.5$ & $1.301 \pm 0.059$ & \cite{Wang2026} \\
  K2-360 b & $7.67 \pm 0.75$ & $1.565 \pm 0.079$ & \cite{Livingston2024} &  TOI-2345 b & $3.49 \pm 0.85$ & $1.504^{+0.047}_{-0.044}$ & \cite{Eschen2025} \\
  K2-38 b & $7.3^{+1.1}_{-1.0}$ & $1.54 \pm 0.14$ & \cite{Toledo-Padron2020} &  TOI-238 b & $3.40^{+0.46}_{-0.45}$ & $1.402^{+0.084}_{-0.086}$ & \cite{SuarezMascareno2024} \\
  Kepler-10 b & $3.24 \pm 0.32$ & $1.47^{+0.03}_{-0.02}$ & \cite{Dumusque2014} &  TOI-238 c & $6.7 \pm 1.1$ & $2.18 \pm 0.18$ & \cite{SuarezMascareno2024} \\
  Kepler-100 b & $4.01 \pm 0.47$ & $1.34 \pm 0.12$ & \cite{Brinkman2025} &  TOI-2431 b & $6.2 \pm 1.6$ & $1.534^{+0.034}_{-0.033}$ & \cite{Tas2026} \\
  Kepler-105 c & $4.60 \pm 0.89$ & $1.31 \pm 0.07$ & \cite{Householder2024} &  TOI-244 b & $2.68 \pm 0.30$ & $1.52 \pm 0.12$ & \cite{Castro-Gonzalez2023} \\
  Kepler-107 c & $10.0 \pm 2.0$ & $1.597 \pm 0.026$ & \cite{Bonomo2019} &  TOI-270 b & $1.48 \pm 0.18$ & $1.28^{+0.05}_{-0.04}$ & \cite{Vaneylen2021} \\
  Kepler-114 c & $1.43^{+0.71}_{-0.47}$ & $1.72 \pm 0.08$ & \cite{Jontof-Hutter2021} &  TOI-270 d & $4.20 \pm 0.16$ & $2.00 \pm 0.05$ & \cite{Vaneylen2021} \\
    Kepler-128 b & $3.79^{+0.76}_{-0.66}$ & $1.421 \pm 0.040$ & \cite{Marcy2014} &  TOI-286 b & $4.53 \pm 0.78$ & $1.42 \pm 0.10$ & \cite{Hobson2024} \\
  Kepler-128 c & $3.38^{+0.67}_{-0.59}$ & $1.521 \pm 0.047$ & \cite{Marcy2014} &  TOI-396 b & $3.55^{+0.94}_{-0.96}$ & $2.004^{+0.045}_{-0.047}$ & \cite{Bonfanti2025} \\
  Kepler-138 c & $2.3^{+0.6}_{-0.5}$ & $1.51 \pm 0.04$ & \cite{Piaulet2023} &  TOI-396 c & $2.24^{+0.13}_{-0.67}$ & $1.979^{+0.054}_{-0.051}$ & \cite{Bonfanti2025} \\
  Kepler-138 d & $2.1^{+0.6}_{-0.7}$ & $1.51 \pm 0.04$ & \cite{Piaulet2023} &  TOI-396 d & $7.1 \pm 1.6$ & $2.001^{+0.063}_{-0.064}$ & \cite{Bonfanti2025} \\
  Kepler-161 b & $2.63^{+0.83}_{-0.61}$ & $2.12 \pm 0.14$ & \cite{Ofir2025} &  TOI-406 c & $2.08^{+0.23}_{-0.22}$ & $1.32 \pm 0.12$ & \cite{Lacedelli2024} \\
  Kepler-161 c & $3.61^{+0.78}_{-1.19}$ & $2.05 \pm 0.15$ & \cite{Ofir2025} &  TOI-406.01 & $6.57^{+1.00}_{-0.90}$ & $2.08^{+0.16}_{-0.15}$ & \cite{Lacedelli2024} \\
  Kepler-1659 b & $9.0 \pm 0.3$ & $1.9 \pm 0.2$ & \cite{Jontof-Hutter2021} &  TOI-431 b & $3.07 \pm 0.35$ & $1.28 \pm 0.04$ & \cite{Osborn2021a} \\
  Kepler-1659 c & $4.6 \pm 0.3$ & $1.9 \pm 0.2$ & \cite{Jontof-Hutter2021} &  TOI-4336 A b & $3.33^{+0.34}_{-0.37}$ & $2.137 \pm 0.080$ & \cite{Parc2026} \\
  Kepler-1705 b & $4.47^{+0.48}_{-0.43}$ & $2.03^{+0.12}_{-0.14}$ & \cite{Leleu2021} &  TOI-4336 A c & $1.55 \pm 0.13$ & $1.251 \pm 0.066$ & \cite{Parc2026} \\
  Kepler-1705 c & $5.42^{+0.61}_{-0.57}$ & $2.05^{+0.14}_{-0.15}$ & \cite{Leleu2021} &  TOI-512 b & $3.57^{+0.53}_{-0.55}$ & $1.54 \pm 0.10$ & \cite{Rodrigues2025} \\
  Kepler-20 b & $9.7 \pm 1.3$ & $1.773^{+0.053}_{-0.030}$ & \cite{Buchhave2016} &  TOI-521 b & $5.3 \pm 1.0$ & $1.98 \pm 0.14$ & \cite{Lacedelli2026} \\
  Kepler-21 b & $7.5 \pm 1.3$ & $1.639^{+0.019}_{-0.015}$ & \cite{Lopez-Morales2016} &  TOI-544 b & $2.89 \pm 0.48$ & $2.018 \pm 0.076$ & \cite{Osborne2024} \\
  Kepler-23 b & $2.56^{+0.43}_{-0.40}$ & $1.638 \pm 0.047$ & \cite{Marcy2014} &  TOI-561 b & $2.02 \pm 0.23$ & $1.397 \pm 0.027$ & \cite{Brinkman2023} \\
  Kepler-238 f & $12.2^{+1.9}_{-1.8}$ & $2.00 \pm 0.09$ & \cite{Ofir2025}, L24 &  TOI-5734 b & $9.1 \pm 2.6$ & $2.10 \pm 0.12$ & \cite{Filomeno2026} \\
  Kepler-30 b & $8.63 \pm 0.42$ & $1.89 \pm 0.13$ & \cite{Jontof-Hutter2022} &  TOI-5788 b & $3.72 \pm 0.94$ & $1.528 \pm 0.075$ & \cite{Lakeland2026} \\
  Kepler-307 c & $3.64 \pm 0.61$ & $2.20 \pm 0.07$ & \cite{Hadden2017} &  TOI-733 b & $5.72^{+0.70}_{-0.68}$ & $1.992^{+0.085}_{-0.090}$ & \cite{Georgieva2023} \\
  Kepler-36 b & $3.83^{+0.11}_{-0.10}$ & $1.498^{+0.061}_{-0.049}$ & \cite{Vissapragada2020} &  TOI-771 b & $2.47^{+0.32}_{-0.31}$ & $1.36 \pm 0.10$ & \cite{Lacedelli2025} \\
  Kepler-406 b & $6.35 \pm 1.40$ & $1.43 \pm 0.03$ & \cite{Marcy2014} &  TOI-776 b & $5.0 \pm 1.6$ & $1.798^{+0.078}_{-0.077}$ & \cite{Luque2021} \\
  Kepler-57 c & $6.86^{+1.52}_{-1.43}$ & $2.196^{+0.067}_{-0.065}$ & \cite{Jontof-Hutter2016} &  TOI-784 b & $9.67^{+0.83}_{-0.82}$ & $1.93^{+0.11}_{-0.09}$ & \cite{Hua2023} \\
  Kepler-60 b & $3.92^{+0.96}_{-0.72}$ & $1.74^{+0.10}_{-0.07}$ &  \cite{Jontof-Hutter2021}, L24   &  TOI-836 b & $4.53^{+0.92}_{-0.86}$ & $1.704 \pm 0.067$ & \cite{Hawthorn2023} \\
  Kepler-60 c & $3.85 \pm 0.81$ & $1.95^{+0.11}_{-0.10}$ & \cite{Jontof-Hutter2021}, L24  &  TOI-912 b & $5.1 \pm 0.5$ & $1.93 \pm 0.13$ & \cite{Lacedelli2026} \\
  Kepler-60 d & $4.60 \pm 0.89$ & $1.76^{+0.14}_{-0.13}$ & \cite{Jontof-Hutter2021}, L24  &  WASP-132 c & $6.26^{+1.84}_{-1.83}$ & $1.841^{+0.094}_{-0.093}$ & \cite{Grieves2025} \\
  Kepler-65 d & $4.14^{+0.79}_{-0.80}$ & $1.52 \pm 0.04$ & \cite{Marcy2014} &  WASP-47 e & $9.0^{+0.6}_{-0.4}$ & $1.83 \pm 0.02$ & \cite{Vanderburg2017} \\
  Kepler-78 b & $1.68 \pm 0.27$ & $1.201 \pm 0.028$ & \cite{Pepe2013} &  Wolf 327 b & $2.53 \pm 0.46$ & $1.24 \pm 0.06$ & \cite{Murgas2024} \\
  Kepler-80 d & $5.95^{+0.65}_{-0.60}$ & $1.309^{+0.036}_{-0.032}$ & \cite{MacDonald2021} &  Wolf 503 b & $6.26^{+0.69}_{-0.70}$ & $2.043 \pm 0.069$ & \cite{Polanski2021} \\
  \hline
\end{tabular}
\end{center}
\end{table}

\begin{table}
  \begin{center}
  \begin{tabular}{|l|ll|l|l|l|}
 \hline
3-body res. & $P_{x}$ min (d) & $P_{x}$ max (d) & $P_{x}$:$P_{c}$ & Num. & $\chi^2_{\rm{min}}$   \\
\hline
$5n_{b}-9n_{c}+4n_{x}$ & 12.5011 & 12.5141 & 4:3 & 163 & 696.4 \\
$5n_{b}-8n_{c}+3n_{x}$ & 14.0640 & 14.0750 & 3:2 & 38 & 705.4  \\
$5n_{b}-7n_{c}+2n_{x}$ & 18.7646 & 18.7827 & 2:1 & 7 & 710.4 \\
\hline
$10n_{b}-19n_{c}+9n_{x}$ & 12.0600 & 12.0633 & 9:7 & 87 & 696.8 \\
\hline
\end{tabular}
\caption{Number of samples and best $\chi^2$ in Case A by resonance (top rows) and Case B (bottom row).}\label{Tab:3bodyResArgs}
\end{center}
\end{table}
\begin{figure*}
\includegraphics[width = 3.5 in]{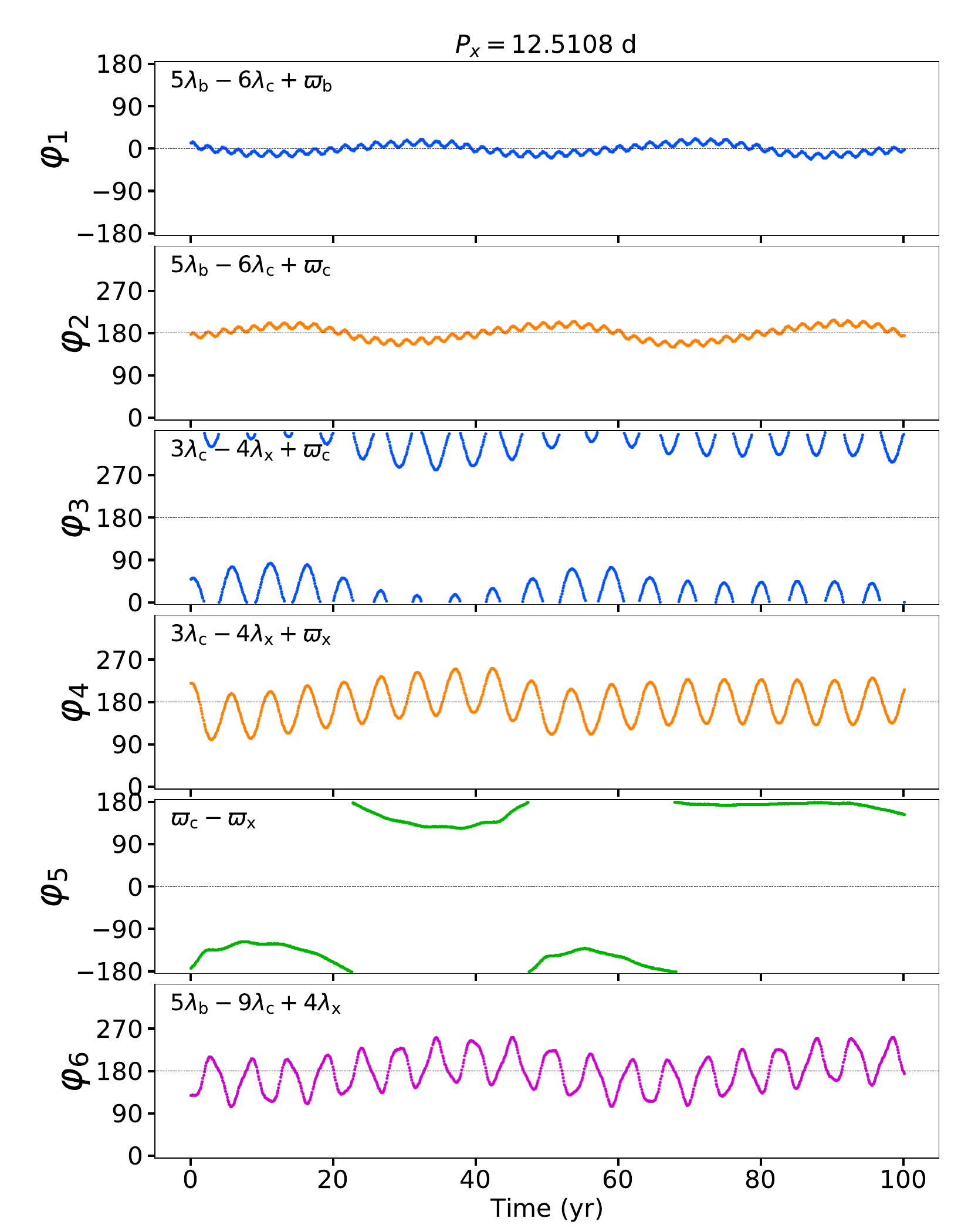}
\includegraphics[width = 3.5 in]{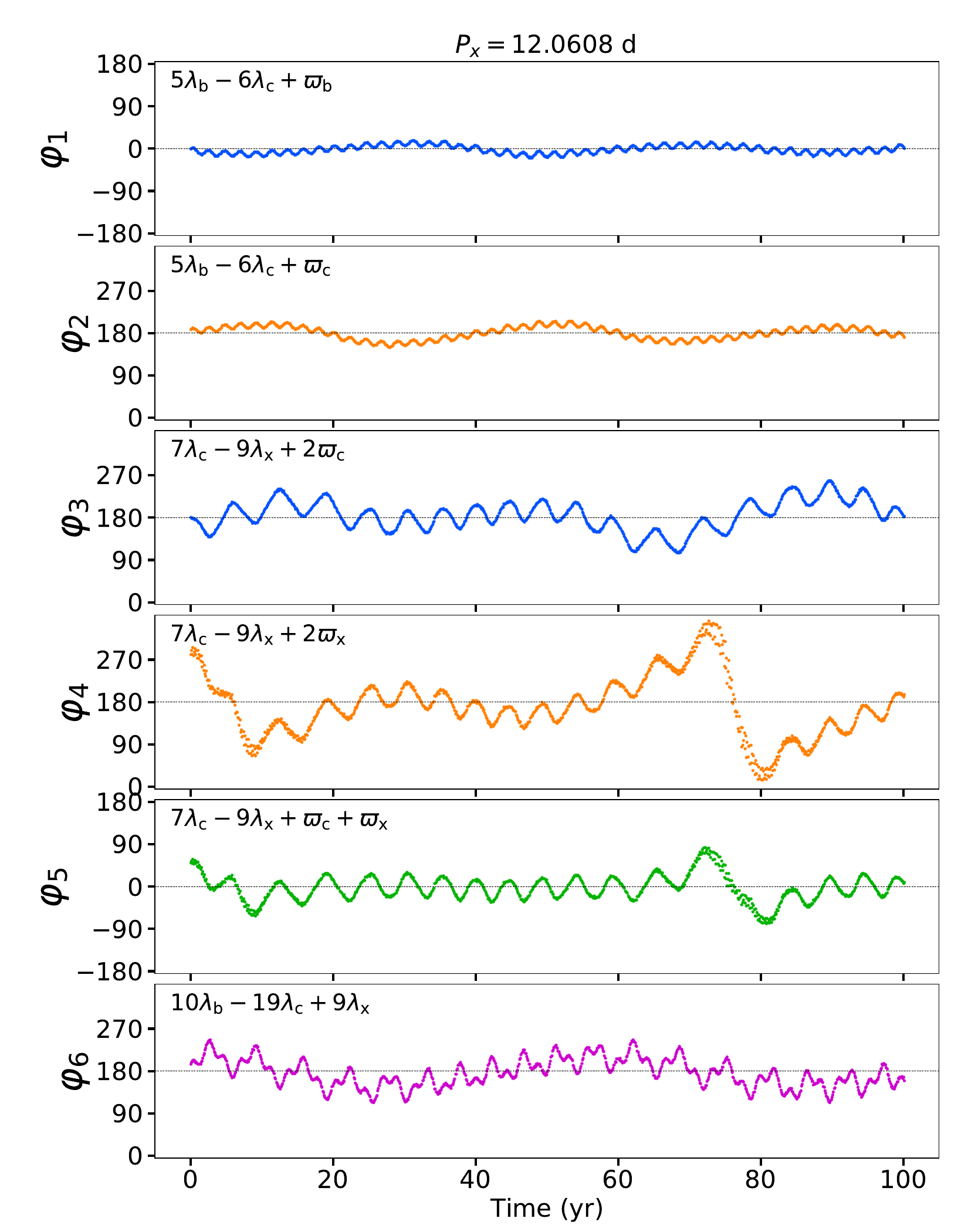}
\caption{Resonant argument plots over 100-year integrations of examples of 3-body resonances from Case A (left) and Case B (right).}
\label{Fig:Phi_t}
\end{figure*}

\clearpage
\bibliography{references}{}
\bibliographystyle{aasjournalv7}
\end{document}